\documentclass[10pt]{article}
\usepackage[letterpaper,margin=1in]{geometry}

\usepackage{amsmath,amsfonts,bm}

\def\eqref#1{equation~\ref{#1}}
\def\1{\bm{1}}

\DeclareMathAlphabet{\mathsfit}{\encodingdefault}{\sfdefault}{m}{sl}
\SetMathAlphabet{\mathsfit}{bold}{\encodingdefault}{\sfdefault}{bx}{n}

\usepackage[utf8]{inputenc} % allow utf-8 input
\usepackage[T1]{fontenc}    % use 8-bit T1 fonts
\usepackage{lmodern}
\usepackage{url}            % simple URL typesetting
\usepackage{booktabs}       % professional-quality tables
\usepackage{amsfonts}       % blackboard math symbols
\usepackage{amsmath}
\usepackage{amssymb}
\usepackage{float}
\usepackage{microtype}      % microtypography
\usepackage{xcolor}
\usepackage{graphicx}
\usepackage{rotating}
\usepackage{natbib}
\usepackage{tikz}
\usepackage[hidelinks]{hyperref}
\usetikzlibrary{shapes.geometric, arrows, positioning}

\tikzstyle{block} = [rectangle, draw, text centered, minimum height=1.5cm, minimum width=3cm]
\tikzstyle{arrow} = [thick,->,>=stealth]
\tikzstyle{shortcut} = [thick,dashed,->,>=stealth]

\title{Flexible Spectral-Normalized Neural Gaussian Process for Dynamic Aperture Prediction}

\author{%
  Yousra El-Bachir\thanks{Corresponding author: yousra.elbachir@epfl.ch} \\
  \small Swiss Data Science Center, ETH Zurich \& EPFL
  \and Frederik Van der Veken \\
  \small CERN, Geneva
  \and Davide di Croce \\
  \small EPFL, Lausanne; CERN, Geneva
  \and Carlo Emilio Montanari \\
  \small CERN, Geneva
  \and Massimo Giovannozzi \\
  \small CERN, Geneva
  \and Ekaterina Krymova\thanks{Corresponding author: ekaterina.krymova@sdsc.ethz.ch} \\
  \small Swiss Data Science Center, ETH Zurich \& EPFL
  \and Tatiana Pieloni \\
  \small EPFL, Lausanne; CERN, Geneva
}

\begin{document}

\maketitle

\begin{abstract}
We address the challenge of scalable uncertainty quantification in large-scale scientific applications, where complex state-of-the-art machine learning methods are often computationally infeasible. Our primary contribution is a simple yet effective empirical Bayes method for automatically tuning the hyperparameters of a flexible, heteroscedastic Spectral-normalized Neural Gaussian Process. This approach retains the expressiveness and uncertainty-awareness of semi-Bayesian neural models while significantly reducing the computational burden by integrating hyperparameter learning directly into the training loop. We demonstrate the practical impact of our method on the task of estimating the dynamic aperture in circular particle accelerators, a fundamental problem in high-energy physics colliders and storage rings, using simulation data from the case of the Large Hadron Collider at CERN. Traditional approaches to DA estimation require extensive particle-tracking simulations, which are prohibitively time-consuming and resource-intensive. Our results show that the proposed method achieves competitive predictive performance and well-calibrated uncertainty estimates at much lower computational cost than state-of-the-art approaches. We stress that, beyond this application, the proposed empirical Bayes framework offers a general solution for training heteroscedastic neural models in situations where manual hyperparameter tuning is impractical. Accordingly, we anticipate that this framework can be applied to other domains that encounter comparable computational limitations.
\end{abstract}

\section{Introduction}
%\subsection{Quick literature review}
Uncertainty in predictive modeling is traditionally categorized into two fundamentally different types: aleatoric and epistemic. Aleatoric uncertainty arises from inherent noise or randomness in the data, such as measurement errors, and is considered irreducible. In contrast, epistemic uncertainty reflects a lack of knowledge due to limited observed data and can, in principle, be reduced by increasing the training set or using more complex models. Accurate disentanglement of these types of uncertainty is particularly important in active learning and safety-critical applications such as health care or autonomous driving \citep{Esteva,huang}. However, recent machine learning literature challenges the assumption that aleatoric and epistemic uncertainties are cleanly separable, showing that they often overlap and interact in practice. Moreover, widely used uncertainty measures produce inconsistent or conflicting results, and current uncertainty quantification methods are poorly calibrated under distributional changes, further motivating the need to develop more robust and coherent approaches and measures; see \cite{smith2024, pmlr-v216-wimmer23a, pmlr-v202-bengs23a, schweighofer2023introducing, kotelevskii2024predictive, postels}. Despite these efforts, reliably disentangling epistemic and aleatoric uncertainty, as well as developing scalable well-calibrated methods, remain open problems. Therefore, in this paper, we take a more pragmatic approach and introduce an efficient method to estimate the total predictive uncertainty directly within the model training process. 

Two of the most commonly used state-of-the-art methods for uncertainty quantification are deep ensembles~(DE) \citep{de} and deep deterministic uncertainty~(DDU) \citep{ddu}. For the classification task, DE involves training multiple deep learning models independently with different random initializations and averaging their predictions. The variance across the ensemble is used to estimate the uncertainty. Although DE improves accuracy, it may still produce overconfident predictions in regions far from the training data, a limitation inherited from the individual base models. Moreover, training multiple deep learning models can be computationally prohibitive in real-world applications, including  scientific ones, making DE impractical in resource-constrained settings. In contrast, DDU estimates predictive uncertainty from a single forward pass of a deterministic neural network trained with residual connections and spectral normalization to mitigate feature collapse. At inference time, DDU estimates the uncertainty from a feature-space density estimator constructed based on Gaussian Discriminant Analysis, fitted separately to each output class. This makes DDU computationally efficient compared to DE, while requiring minimal architectural modifications. A more advanced state-of-the-art method is the Spectral-normalized Neural Gaussian Process (SNGP) \citep{sngp}, along with its heteroscedastic extension \citep{hetsngp}. This semi-Bayesian approach combines the expressive predictive capabilities of deep neural networks with the principled uncertainty estimation of Gaussian Processes (GPs), and produces a better calibrated uncertainty under distributional shifts compared to several competing methods~\citep{postels}. By appending a GP layer to a residual network with spectrally normalized weights, SNGP becomes sensitive to the distance between training and test points, which is particularly advantageous for out-of-distribution detection and helps mitigate the overconfidence observed in DE. However, SNGP’s reliance on manual hyperparameter tuning makes it less computationally attractive in practice. We propose to improve the flexibility of SNGP and integrate its hyperparameter tuning directly into the training process, thereby improving usability without compromising performance. We motivate the development of this method through a critical problem in the design of circular particle accelerators for high-energy physics, namely, the prediction of dynamic aperture (DA) and possibly particle loss rates; see, e.g. ~\cite{schenk:ipac21-tupab216, MDPI_ML, da_gp, da_svm}, based on the simulation data from the CERN Large Hadron Collider (LHC)~\cite{LHCDR}.% and its luminosity upgrade (HL-LHC)~\cite{Aberle:2749422}.

%\subsection{Motivation}
The DA is a fundamental concept in accelerator physics, representing the extent of the region of phase space in which particle trajectories remain bounded over a predefined number of revolutions around a circular accelerator. Particles starting outside of this region are ultimately lost, resulting in beam degradation and reduced operational efficiency, making the DA a key indicator of long-term beam stability. Therefore, accurate prediction of the DA is critical for understanding nonlinear beam dynamics and for optimizing both the performance and the safe operation of modern colliders and storage rings; see, e.g. ~\cite{PhysRevE.53.4067,dynap1,invlog,da_and_losses,giovannozzi:2026}. For the LHC, determining the DA traditionally involves scanning a high-dimensional parameter space through particle-tracking simulations, often guided by expert accelerator physics knowledge and heuristic strategies. This process is both computationally intensive and time-consuming, making DA studies challenging. In this paper, in contrast to direct DA prediction \cite{computers14070287, Di_Croce_2024}, we introduce an efficient and automated approach to predicting DA by distinguishing the stable region from the rest of the phase space. Moreover, we produce reliable uncertainty estimates that can guide tracking efforts toward regions where predictions are uncertain.

\begin{figure}[H]
\begin{minipage}[H]{.33\textwidth}
\centering
\includegraphics[width=\textwidth]{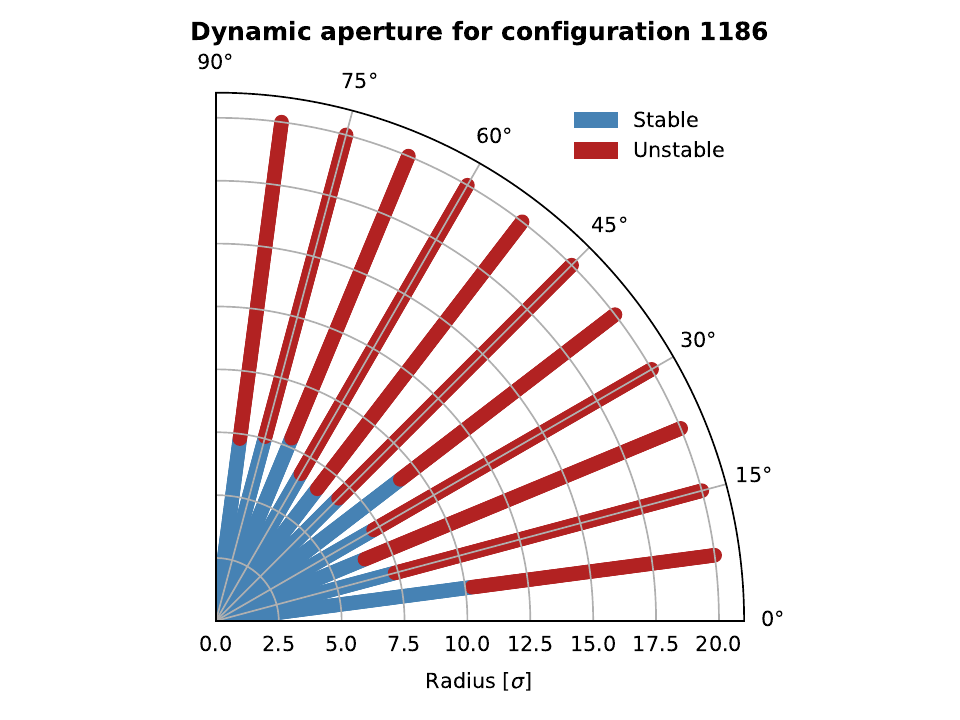}
\end{minipage}
\hfill
\begin{minipage}[H]{.32\textwidth}
\centering
\includegraphics[width=\textwidth]{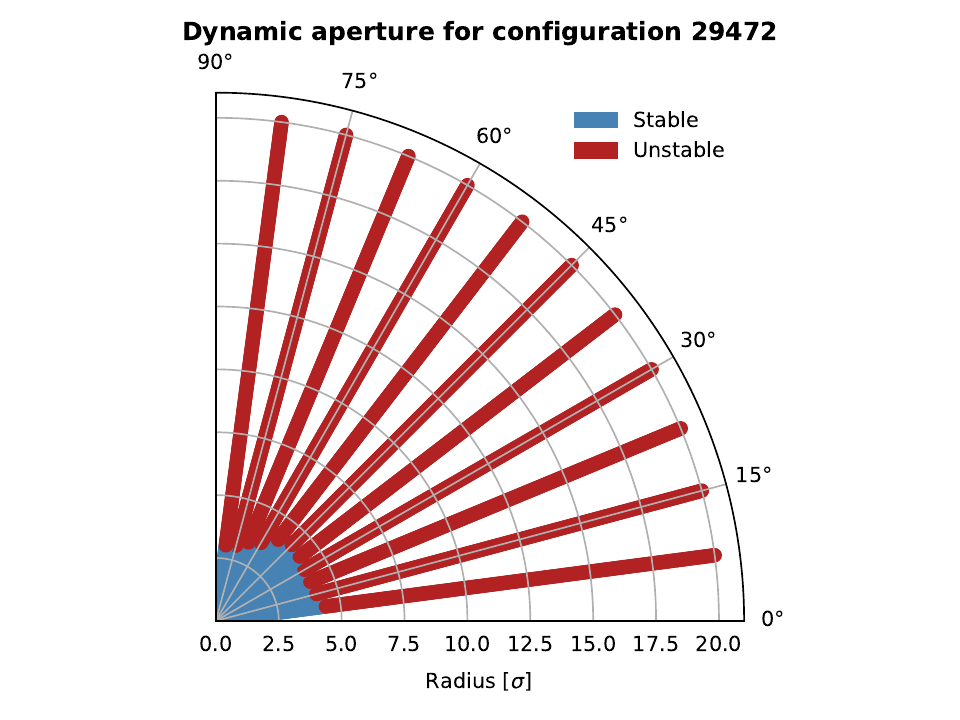}
\end{minipage}  
\hfill
\begin{minipage}[H]{.33\textwidth}
\centering
\includegraphics[width=\textwidth]{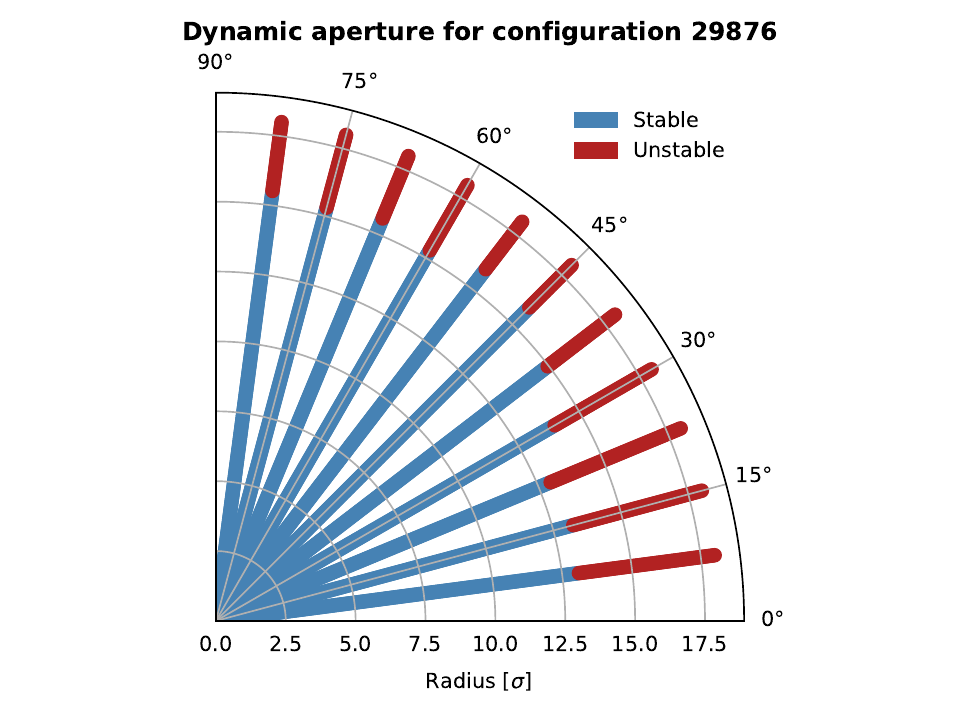}
\end{minipage} 
\caption{Stability regions derived from particle-tracking simulations of the LHC under three different accelerator configurations, with particles initialized at varying radial positions and angular orientations.} \label{fig:original}
\end{figure}

%\subsection{The problem}
Particle-tracking simulations for the LHC, such as SixTrack~\cite{demaria:ipac19-wepts043} with MAD-X\footnote{\url{https://abpcomputing.web.cern.ch/codes/codes_pages/MadX/}}, enable the identification of three distinct spatial regions within a particle accelerator. Particles that complete the maximum number of observable revolutions define the stable region of the phase space. In contrast, the unstable region corresponds to the coordinates where the particles are rapidly lost, failing to complete a significant number of revolutions. Between these two extremes lies a chaotic boundary region, where particles survive a substantial---but submaximal---number of revolutions before being lost. Figure~\ref{fig:original} illustrates the DA, corresponding to the stable region, for three randomly selected accelerator configurations. Each configuration is defined by a unique combination of the values of the input control parameters used in the simulator. For simplicity, the chaotic boundary and the unstable region are merged, as the primary focus is on accurately predicting the stable region. These examples already highlight several challenges associated with modeling the DA. First, the stable region has a complex and nonlinear structure that cannot be captured by simple parametric models, motivating the use of data-driven approaches. Second, the inherently noisy and ambiguous structure of the chaotic boundary makes the classification problem difficult. Third, the dataset represents simulations for millions of particles given hundreds of control variables, making the modeling task computationally intensive. The objective is therefore to develop a flexible, data-driven classification model for predicting the DA, designed for efficient training with minimal dependence on hyperparameter tuning. Crucially, the model's uncertainty estimates should be well-calibrated: they should be high only in regions where predictions are likely to be unreliable.

The paper is organized as follows. Section~\ref{sec:modeling} extends the standard SNGP architecture to accommodate more flexible settings and describes the residual network architecture tailored to the tabular data used in our study. Section~\ref{sec:ebopt} introduces an empirical Bayes framework that enables automatic integration of hyperparameter tuning into the training process. Section~\ref{sec:results} presents the empirical results, and Section~\ref{sec:discussion} closes the article with a discussion of the findings and potential future research directions.

\section{The flexible auto-hetSNGP model}\label{sec:modeling}
\begin{figure}[H]
\begin{minipage}[b]{.5\textwidth}
\centering
\includegraphics[width=0.7\textwidth]{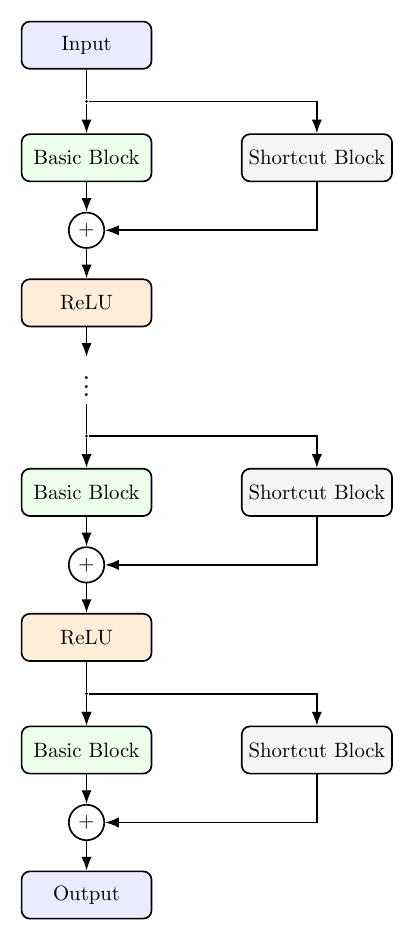}
\end{minipage}
\hfill
\begin{minipage}[b]{.5\textwidth}
\centering
\includegraphics[width=0.8\textwidth]{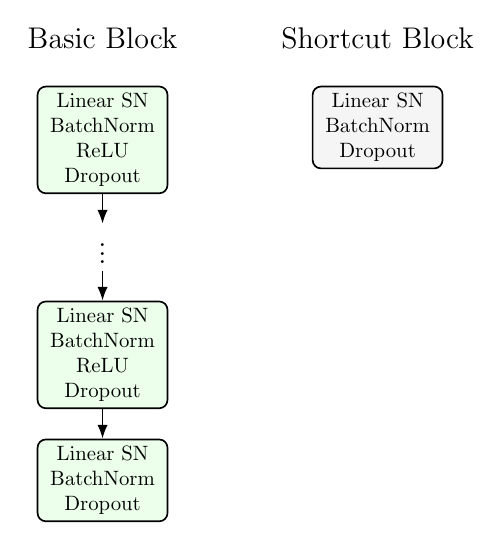}
\end{minipage}  
\caption{Residual network with skip connections and Spectral Normalization (SN).}\label{fig:res}
\end{figure}

The heteroscedastic Spectral-normalized Neural Gaussian Process version (hetSNGP) introduced by \citet{hetsngp} extends the original SNGP framework of \citet{sngp} by incorporating the heteroscedastic model of \citet{heteroscedastic} using a latent variable drawn from an approximate Gaussian Process, as follows. Let $\{(\boldsymbol{x}_i,y_i)\}_{i=1}^n$ be  $n$ independent pairs of input vectors $\boldsymbol{x}_i \in \mathbb{R}^d$ and outputs $y_i \in \{1,\dots,K\}$. Each label $y_i$ is modeled by the softmax distribution over a latent random vector $\boldsymbol{u}(\boldsymbol{x}_i) = \{u_1(\boldsymbol{x}_i), \ldots, u_K(\boldsymbol{x}_i)\} \in \mathbb{R}^{1\times K}$,  
\begin{eqnarray}
P\{y_i = c \mid \boldsymbol{u}(\boldsymbol{x}_i)\} = \text{Softmax}_{\tau}\{\boldsymbol{u}(\boldsymbol{x}_i)\} = \frac{\exp\{u_c(\boldsymbol{x}_i)/\tau\}}{\sum_{k=1}^K \exp \{u_k(\boldsymbol{x}_i)/\tau\}} \label{eq:m1}
\end{eqnarray}
for $c=1, \ldots, K$, where $\tau$ is a temperature parameter. Each component $u_c(\boldsymbol{x}_i)$ is modeled with a mean-variance decomposition
\begin{eqnarray}
u_c(\boldsymbol{x}_i) &=& \mu_c(\boldsymbol{x}_i) + \epsilon_c(\boldsymbol{x}_i), \label{eq:meanvardec}
\end{eqnarray}
where $\mu_c$ denotes the predictive mean, and $\epsilon_c$ represents a zero-mean stochastic class-specific deviation capturing heteroscedastic uncertainty. The predictive mean is modeled using Bayesian linear regression,
\begin{eqnarray}
\mu_c(\boldsymbol{x}_i) &=& \Phi(\boldsymbol{x}_i)\boldsymbol{\beta}_c, \quad \boldsymbol{\beta}_c \sim \mathcal{N}(\boldsymbol{0}, \boldsymbol{I}), \label{eq:bayes}
\end{eqnarray}
where $\boldsymbol{\beta}_c \in \mathbb{R}^{D_L \times 1}$ is a class-specific weight vector, and $\Phi(\boldsymbol{x}_i) \in \mathbb{R}^{1 \times D_L}$ is a random feature embedding constructed as follows. We first adapt the ResNet-like architecture that the hetSNGP model was originally trained on to our tabular data problem. The input $\boldsymbol{x}_i$ is first passed through a residual neural network illustrated in Figure~\ref{fig:res}, which comprises a sequence of residual blocks, each consisting of a basic block paired with an associated shortcut block. The shortcut block has a single linear layer and serves two key purposes: it ensures dimensional compatibility between the input and the output of the main path, and it preserves feature identity, which facilitates stable gradient flow during training. Each basic block is implemented as a multilayer perceptron, where each linear layer has spectrally normalized (SN) weights, followed by batch normalization, a ReLU activation, and dropout. Spectral normalization is applied to preserve the Lipschitz continuity of the forward pass and to maintain geometric consistency in the feature space.
The resulting hidden representation $\boldsymbol{h}(\boldsymbol{x}_i) \in \mathbb{R}^{D_{L-1} \times 1}$ is then passed through a GP output layer approximated by Random Fourier Features (RFF)~\citep{RFF}
\begin{eqnarray*}
\Phi(\boldsymbol{x}_i) &=& \sqrt{2/D_{L}} \cos \{\boldsymbol{W}\boldsymbol{h}(\boldsymbol{x}_i) + \boldsymbol{b} \}^T,
\end{eqnarray*}
with $\boldsymbol{W} \in \mathbb{R}^{D_L \times D_{L-1}}$ and $\boldsymbol{b} \in \mathbb{R}^{D_L \times 1}$ drawn from
\begin{eqnarray*}
W_{ij} &\sim &  \mathcal{N}(0, 1), \quad b_i \sim \mathcal{U}(0, 2\pi),
\end{eqnarray*}
and kept fixed during training. This RFF-based layer approximates a stationary GP kernel and ensures distance-awareness of the network output, which improves sensitivity to out-of-distribution samples, while the Bayesian formulation in~\ref{eq:bayes} provides a principled way to encode prior uncertainty over the mean function. 

The stochastic component $\epsilon_{c}$ in~\ref{eq:meanvardec} models heteroscedasticity through the following combination of class-specific and shared noise terms
\begin{eqnarray}
 \epsilon_c(\boldsymbol{x}_i) &=& d_c(\boldsymbol{x}_i) \kappa_{c} + \boldsymbol{v}_{c}(\boldsymbol{x}_i) \boldsymbol{\gamma}, \label{eq:eps}\\
     \kappa_{c} &\sim &  \mathcal{N}(0, 1), \quad \boldsymbol{\gamma} \sim \mathcal{N}(\boldsymbol{0}, \boldsymbol{I}), \label{eq:ident}
\end{eqnarray}
where $\kappa_c$ is a class-specific noise term, $\boldsymbol{\gamma} \in \mathbb{R}^{R \times 1}$ is a global latent noise vector of dimension $R \ll K$, and $d_c(\boldsymbol{x}_i) \in \mathbb{R}$ and $\boldsymbol{v}_{c}(\boldsymbol{x}_i)\in \mathbb{R}^{1 \times R}$ are input-dependent coefficients obtained as linear projections of the hidden representation $\boldsymbol{h}(\boldsymbol{x}_i)$,
\begin{eqnarray}
\boldsymbol{d}(\boldsymbol{x}_i) &=&  \boldsymbol{W}_d\boldsymbol{h}(\boldsymbol{x}_i) + \boldsymbol{b}_d, \label{eq:dd} \\
\boldsymbol{V}(\boldsymbol{x}_i) &=& {\rm Reshape}\{\boldsymbol{W}_V\boldsymbol{h}(\boldsymbol{x}_i) + \boldsymbol{b}_V, K, R\}, \label{eq:VV}
\end{eqnarray}
where $\boldsymbol{d}(\boldsymbol{x}_i) \in \mathbb{R}^{1\times K}$ and $\boldsymbol{V}(\boldsymbol{x}_i)\in  \mathbb{R}^{K\times R}$ have learnable parameters $\boldsymbol{W}_d \in \mathbb{R}^{K \times D_{L-1}}$, $\boldsymbol{b}_d \in \mathbb{R}^{K \times 1}$, $\boldsymbol{W}_V \in \mathbb{R}^{KR \times D_{L-1}}$ and $\boldsymbol{b}_V \in \mathbb{R}^{KR \times 1}$. The condition $R \ll K$ imposes a low-rank structure, and the inclusion of $\boldsymbol{d}(\boldsymbol{x}_i)$ ensures that the covariance matrix of the latent variable $\boldsymbol{u}(\boldsymbol{x}_i)$ remains strictly positive definite. %The term $d_c(\boldsymbol{x}_i) \kappa_c$ in~\ref{eq:eps} introduces independent noise per class, whereas the low-rank latent factor $\boldsymbol{v}_c(\boldsymbol{x}_i) \boldsymbol{\gamma}$ models correlations across classes by sharing the global noise vector $\boldsymbol{\gamma}$. 
The structured noise formulation in~\ref{eq:eps} therefore enables the hetSNGP model to capture both independent class-specific noise through the term $d_c(\boldsymbol{x}_i) \kappa_c$ and correlations across classes by sharing the global noise~$\boldsymbol{\gamma}$. %through the low-rank latent factor $\boldsymbol{v}_c(\boldsymbol{x}_i) \boldsymbol{\gamma}$. 

%However, this expressiveness comes at the cost of limited flexibility, as the model relies on a single temperature parameter $\tau$ for calibration. 
The hetSNGP model in~\ref{eq:m1} introduces a global temperature parameter~$\tau$, which in the original article \citep{hetsngp} was kept fixed during training and prediction, showing that its value did not significantly affect the quality of uncertainty calibration in an ablation study in the ImageNet dataset. In general, the temperature parameter can be tuned post hoc to control the calibration of predictive uncertainty \citep{guo2017calibration}, at the cost of additional computational overhead that we propose to relieve as follows. On setting $\tilde{u}_c(\boldsymbol{x}_i) = u_c(\boldsymbol{x}_i)/\tau$ and rewriting the temperature-scaled softmax in~\ref{eq:m1}, we get
\begin{eqnarray}
P\{y_i = c \mid \boldsymbol{u}(\boldsymbol{x}_i)\} &= & \text{Softmax}\{\boldsymbol{\tilde u}(\boldsymbol{x}_i)\}, \nonumber \\
\tilde{u}_c(\boldsymbol{x}_i) &=& \Phi(\boldsymbol{x}_i)\boldsymbol{\tilde \beta}_c + d_c(\boldsymbol{x}_i) \tilde{\kappa}_{c} + \boldsymbol{v}_{c}(\boldsymbol{x}_i) \boldsymbol{\tilde \gamma}, \nonumber \\
    \boldsymbol{\tilde \beta}_c &\sim & \mathcal{N}(\boldsymbol{0}, 1/\tau^2 \boldsymbol{I}), \quad \tilde{\kappa}_{c} \sim  \mathcal{N}(0, 1/\tau^2), \quad \boldsymbol{\tilde \gamma} \sim  \mathcal{N}(0, 1/\tau^2 \boldsymbol{I}), \label{eq:scaling}
\end{eqnarray}
where the factor $1/\tau^2$ in $\tilde{\kappa}_c$ and in $\boldsymbol{\tilde{\gamma}}$ can be absorbed into the learnable parameters $\boldsymbol{W}_d$, $\boldsymbol{b}_d$, $\boldsymbol{W}_V$ and $\boldsymbol{b}_V$ in~\ref{eq:dd}–\ref{eq:VV}. The explicit scaling in~\ref{eq:scaling} therefore becomes redundant relative to the reduced form in~\ref{eq:ident}, but this simplification is only valid when $\boldsymbol{d}(\boldsymbol{x}_i)$ and $\boldsymbol{V}(\boldsymbol{x}_i)$ are linear parameterizations as in~\ref{eq:dd}–\ref{eq:VV}. However, we maintain the scaling in the definitions of $\tilde{\kappa}_c$ and $\boldsymbol{\tilde{\gamma}}$ throughout the remainder of the paper to allow seamless extension to settings where explicit control over the variance is essential and to highlight the generality of the proposed optimization framework. We also improve the original hetSNGP model's flexibility by learning class-dependent variance terms through the following reformulation
\begin{eqnarray}
P\{y_i = c \mid \boldsymbol{u}(\boldsymbol{x}_i)\} &= & \text{Softmax}_{\tau =1}\{\boldsymbol{u}(\boldsymbol{x}_i)\}, \label{eq:main_model}\\
u_c(\boldsymbol{x}_i) &=& \Phi(\boldsymbol{x}_i)\boldsymbol{\beta}_c + d_c(\boldsymbol{x}_i) \kappa_{c} + \boldsymbol{v}_{c}(\boldsymbol{x}_i) \boldsymbol{\gamma}, \label{eq:u}\\
    \boldsymbol{\beta}_c &\sim & \mathcal{N}(\boldsymbol{0}, \sigma^2_{b_c} \boldsymbol{I}), \quad \kappa_{c} \sim  \mathcal{N}(0, \sigma^2_{\kappa_c}), \quad \gamma_{j} \sim  \mathcal{N}(0, \sigma^2_{\gamma_j}), \label{eq:priors}
\end{eqnarray}
and we tune additionally introduced parameters by adopting an empirical Bayes approach that integrates hyperparameter estimation directly into the training process, thereby enhancing scalability. By introducing class-specific variance for $\boldsymbol{\beta}_c$, we additionally enable the model to automatically learn the appropriate levels of regularization for each class. This flexibility allows the model to account for class-specific signal strength and potential class imbalance, improving predictive performance and uncertainty calibration. We now describe the optimization procedure. 

\section{Empirical Bayes optimization}\label{sec:ebopt}
Let $\boldsymbol{\beta} = (\boldsymbol{\beta}_1, \ldots, \boldsymbol{\beta}_K)$ and $\boldsymbol{\kappa} = (\kappa_1, \ldots, \kappa_K)$ denote the model weights, and let $\boldsymbol{\tau}_{b} = \{\log(\sigma^2_{b_1}), \ldots, \log(\sigma^2_{b_K})\}$, $\boldsymbol{\tau}_{\kappa} = \{\log(\sigma^2_{\kappa_1}), \ldots, \log(\sigma^2_{\kappa_K})\}$ and $\boldsymbol{\tau}_{\gamma} = \{\log(\sigma^2_{\gamma_1}), \ldots, \log(\sigma^2_{\gamma_R})\}$ denote the corresponding log-variances. Learning logarithms instead of original variances mitigates potential numerical instabilities by performing the optimization in an unconstrained parameter space. The optimal variances are then recovered by exponentiation of the learned log-variance values. We divide the model parameters into two groups, a vector of deterministic variables denoted by $\boldsymbol{\psi} = \left(\boldsymbol{\tau}_{b}, \boldsymbol{\tau}_{\kappa}, \boldsymbol{\tau}_{\gamma}, \boldsymbol{h}, \boldsymbol{W}_d, \boldsymbol{b}_d, \boldsymbol{W}_V, \boldsymbol{b}_V \right)$, which we learn during the training phase, and a vector of stochastic variables denoted by $\boldsymbol{\theta} = \left(\boldsymbol{\beta}, \boldsymbol{\kappa}, \boldsymbol{\gamma} \right)$, which we use for posterior sampling during prediction. We train the auto-hetSNGP model in~\ref{eq:main_model}--\ref{eq:priors} by maximizing the log-marginal likelihood with respect to the deterministic parameters to obtain the optimal estimate $\boldsymbol{\hat{\psi}}$, which we hold fixed at inference time. We then compute the predictions by sampling from the posterior distribution of $\boldsymbol{\theta}$ given $\boldsymbol{y}$ and fixed $\boldsymbol{\hat{\psi}}$ using the Laplace method. Overall, the empirical Bayes method for the flexible model amounts to a double optimization, which we now develop.

\subsection{Double maximization}
\paragraph{The marginal likelihood based loss function for $\boldsymbol{\psi}$.} We denote the prior distribution of $\boldsymbol{\theta}$ for a fixed~$\boldsymbol{\psi}$ by 
\begin{eqnarray*}
    p(\boldsymbol{\theta}; \boldsymbol{\psi})= p(\boldsymbol{\beta}; \boldsymbol{\sigma}^2_{b}) \ p(\boldsymbol{\kappa}; \boldsymbol{\sigma}^2_{\kappa}) \ p(\boldsymbol{\gamma}; \boldsymbol{\sigma}^2_{\gamma}),
    \end{eqnarray*}
as a product of individual priors in~\ref{eq:priors}. The marginal likelihood is hence
\begin{eqnarray*}
p(\boldsymbol{y};\boldsymbol{x}, \boldsymbol{\psi}) &=& \int p\left(\boldsymbol{y} \mid \boldsymbol{\theta}; \boldsymbol{x}, \boldsymbol{\psi} \right) p(\boldsymbol{\theta}; \boldsymbol{\psi}) \, \mathrm{d}\boldsymbol{\theta}\\
&=& \text{E}_{\boldsymbol{\theta} \sim p(\boldsymbol{\theta}; \boldsymbol{\psi})} \left[\exp\left\{\sum_{i=1}^n \log p\left(y_i \mid \boldsymbol{\theta}; \boldsymbol{x}_i, \boldsymbol{\psi}\right) \right\} \right]\\
& \approx & \cfrac{1}{S}\sum_{s=1}^S \exp\left[\sum_{i=1}^n \log \text{Softmax}\left\{\boldsymbol{u}_{y_i}^{(s)}(\boldsymbol{x}_i)\right\} \right],
\end{eqnarray*}
where the approximation follows from $S$ Monte-Carlo samples, each of which is generated by
\begin{eqnarray*}
\boldsymbol{u}_{y_i}^{(s)}(\boldsymbol{x}_i) &=& \Phi(\boldsymbol{x}_i)\boldsymbol{\beta}_{y_i}^{(s)} + d_{y_i}(\boldsymbol{x}_i) \kappa_{y_i}^{(s)} + \boldsymbol{v}_{y_i}(\boldsymbol{x}_i) \boldsymbol{\gamma}^{(s)},
\end{eqnarray*}
with samples drawn from the respective priors
\begin{eqnarray*}
    \boldsymbol{\beta}^{(s)}_{y_i} &\sim & \mathcal{N}(\boldsymbol{0}, \sigma^2_{b_{y_i}} \boldsymbol{I}), \quad \kappa^{(s)}_{y_i} \sim  \mathcal{N}(0, \sigma^2_{\kappa_{y_i}}), \quad \gamma^{(s)}_{j} \sim  \mathcal{N}(0, \sigma^2_{\gamma_j}).
\end{eqnarray*}
At training time, the optimal deterministic parameters $\boldsymbol{\hat \psi}$ are obtained by minimizing $-\log p(\boldsymbol{y};\boldsymbol{x}, \boldsymbol{\psi})$ using stochastic gradient descent or one of its variants. At inference time, we need to sample from the posterior distribution $p(\boldsymbol{\theta} \mid \boldsymbol{y}; \boldsymbol{x}, \boldsymbol{\hat{\psi}})$, which we approximate by a Gaussian distribution whose expectation and covariance matrix are learned from the maximum penalized likelihood estimator as we shall now see. 

\paragraph{The penalized likelihood based loss function for $\boldsymbol{\theta}$.} We denote the log-penalized likelihood of $\boldsymbol{\theta}$ for fixed $\boldsymbol{\hat \psi}$ by 
$$\ell_{\rm P}(\boldsymbol{\theta}; \boldsymbol{y}, \boldsymbol{x}, \boldsymbol{\hat \psi}) = \log \{ p(\boldsymbol{y} \mid \boldsymbol{\theta}; \boldsymbol{x}, \boldsymbol{\hat \psi}) \ p(\boldsymbol{\theta}; \boldsymbol{\hat \psi})\}.$$ Using the hierarchical model in~\ref{eq:main_model}--\ref{eq:priors}, we obtain
\begin{small}
\begin{eqnarray}
\ell_{\rm P}(\boldsymbol{\theta}; \boldsymbol{y}, \boldsymbol{x}, \boldsymbol{\hat \psi})
& \equiv & \sum_{i=1}^n \log \text{Softmax}\left\{\boldsymbol{u}_{y_i}(\boldsymbol{x}_i)\right\} - \cfrac{1}{2} \left\{\sum_{c=1}^K \left (\cfrac{1}{\hat \sigma_{b_c}^2}\ \| \boldsymbol{\beta}_c \|_2^2  + \cfrac{1}{\hat \sigma_{\kappa_c}^2}\ \kappa^2_{c} \right ) + \sum_{r=1}^R
\cfrac{1}{\hat \sigma_{\gamma_r}^2} \ \gamma^2_{r} \right \}, \label{eq:lp}
\end{eqnarray}
\end{small}
where
\begin{eqnarray*}
\boldsymbol{u}_{y_i}(\boldsymbol{x}_i) &=& \hat{\Phi}(\boldsymbol{x}_i)\boldsymbol{\beta}_{y_i} + \hat{d}_{y_i}(\boldsymbol{x}_i) \kappa_{y_i} + \boldsymbol{\hat{v}}_{y_i}(\boldsymbol{x}_i) \boldsymbol{\gamma},
\end{eqnarray*}
and the weights are now considered learnable deterministic parameters.
\noindent Let $\boldsymbol{H}_{\rm P}(\boldsymbol{\theta}; \boldsymbol{y}, \boldsymbol{x}, \boldsymbol{\hat \psi})$ denote the hessian of the negative log-penalized likelihood, and let $\boldsymbol{\hat{\theta}_{\hat{\psi}}} = \arg \max_{\boldsymbol{\theta}} \ell_{\rm P}(\boldsymbol{\theta}; \boldsymbol{y}, \boldsymbol{x}, \boldsymbol{\hat \psi})$ denote
the maximizer. The Laplace approximation to the posterior likelihood leads to 
\begin{eqnarray}
p(\boldsymbol{\theta} \mid \boldsymbol{y}; \boldsymbol{x}, \boldsymbol{\hat \psi}) &=& \cfrac{p(\boldsymbol{y} \mid \boldsymbol{\theta}; \boldsymbol{x}, \boldsymbol{\hat \psi}) \ p(\boldsymbol{\theta}; \boldsymbol{ \hat{\psi}}) }{p(\boldsymbol{y}; \boldsymbol{x}, \boldsymbol{ \hat{\psi}})} \nonumber\\
&\propto & \exp \left\{\ell_{\rm P}(\boldsymbol{\theta}; \boldsymbol{y}, \boldsymbol{x}, \boldsymbol{\hat \psi}) \right\} \nonumber \\
&\approx & \mathcal{N}\left\{ \boldsymbol{\hat{\theta}_{\hat{\psi}}};  \boldsymbol{H}^{-1}_{\rm P}(\boldsymbol{\hat{\theta}_{\hat{\psi}}}; \boldsymbol{y}, \boldsymbol{x}, \boldsymbol{\hat \psi}) \right\}. \label{eq:post}
\end{eqnarray}
Once the log-penalized likelihood in~\ref{eq:lp} is maximized to obtain $\boldsymbol{\hat{\theta}_{\hat{\psi}}}$, sampling $\boldsymbol{\theta}$ from its posterior distribution involves evaluating $\boldsymbol{H}_{\rm P}(\boldsymbol{\theta}; \boldsymbol{y}, \boldsymbol{x}, \boldsymbol{\hat \psi})$ at $\boldsymbol{\theta} = \boldsymbol{\hat{\theta}_{\hat{\psi}}}$, and drawing samples from the corresponding Gaussian distribution~\ref{eq:post}. 

The dissociation between the deterministic $\boldsymbol{\psi}$ and the stochastic $\boldsymbol{\theta}$ significantly reduces computational overhead, as the expensive training of the neural network is performed only once at the minimization of the negative log-marginal likelihood. Similarly, the optimization of the log-penalized likelihood on the training set is performed only once at inference time for evaluating predictive accuracy. Crucially, the input-dependent terms $\boldsymbol{\Phi}(\boldsymbol{x})$, $d(\boldsymbol{x})$ and $\boldsymbol{v}(\boldsymbol{x})$ of the latent variable $\boldsymbol{u}(\boldsymbol{x})$ are computed from a single forward pass through the neural network and are shared accross the evaluations of both the marginal and penalized likelihoods. We now detail the minimization of the negative log-penalized likelihood in~\ref{eq:lp} using the Newton--Raphson algorithm, which is feasible in our setting given the low dimensionality of $\boldsymbol{\theta}$. 

\paragraph{Stable implementation of the Newton--Raphson minimizer for the log-penalized likelihood.} 
We develop a robust optimization algorithm using preconditioning to improve numerical stability and convergence even in cases where the Hessian matrix may be ill-conditioned.

Let $\boldsymbol{G}_{\rm P} (\boldsymbol{\theta}; \boldsymbol{\hat{\psi}})$ and $\boldsymbol{H}_{\rm P} (\boldsymbol{\theta}; \boldsymbol{\hat{\psi}})$ denote the gradient vector and Hessian matrix of the negative log-penalized likelihood, respectively. For notational simplicity, we omit the explicit dependence on $\boldsymbol{y}$ and $\boldsymbol{x}$. Given an intermediate update $\boldsymbol{\theta}^{(k)}$, one iteration of the minimization of $-\ell_{\rm P}$ proceeds as follows: 
\begin{enumerate}
\item compute a robust Newton--Raphson step that guarantees a stable descent direction:
\begin{enumerate}
\item extract a preconditioner $\boldsymbol{P} = |\boldsymbol{H}_{ii}|^{-1/2}$ from the diagonal of $\boldsymbol{H}_{\rm P}$, and form the preconditioned Hessian $\boldsymbol{\tilde{H}}_{\rm P} = \boldsymbol{P} \boldsymbol{H}_{\rm P} \boldsymbol{P}$; 
\item enforce positive definiteness of $\boldsymbol{\tilde{H}}_{\rm P}$ by performing an eigendecomposition $\boldsymbol{\tilde{H}}_{\rm P} = \boldsymbol{V} \boldsymbol{\Lambda} \boldsymbol{V}^T$, and replacing any non-positive or near-zero eigenvalues in $\boldsymbol{\Lambda}$ with the smallest positive eigenvalue;
\item compute the Newton--Raphson update direction $\boldsymbol{\Delta}^{(k)}$ as
\begin{eqnarray}
\boldsymbol{\Delta}^{(k)} &=& \boldsymbol{H}^{-1}_{\rm P}(\boldsymbol{\theta}^{(k)}; \boldsymbol{\hat{\psi}}) \ \boldsymbol{G}_{\rm P} (\boldsymbol{\theta}^{(k)}; \boldsymbol{\hat{\psi}}) \nonumber \\
&=& \boldsymbol{P} \boldsymbol{V} \boldsymbol{\Lambda}^{-1} \boldsymbol{V}^T \boldsymbol{P} \ \boldsymbol{G}_{\rm P} (\boldsymbol{\theta}^{(k)}; \boldsymbol{\hat{\psi}}), \label{eq:prod}
\end{eqnarray}
where the matrix-vector product on the right hand-side of~\ref{eq:prod} is efficiently evaluated from right to left;
\end{enumerate}
\item compute a candidate
\begin{eqnarray}
\boldsymbol{\theta^*} &=& \boldsymbol{\theta}^{(k)} - \delta \boldsymbol{\Delta}^{(k)},\label{eq:NR}
\end{eqnarray}
with initial learning rate $\delta=1$, 
\item tune the learning rate by repeatedly halving $\delta$ and evaluating \ref{eq:NR} until $-\ell_{\rm P}(\boldsymbol{\theta^*}; \boldsymbol{\hat{\psi}})< -\ell_{\rm P}(\boldsymbol{\theta}^{(k)};\boldsymbol{\hat{\psi}})$;
\item updates $\boldsymbol{\theta}^{(k)}$ with $\boldsymbol{\theta^*}$.
\end{enumerate}
At convergence, the last update $\boldsymbol{\theta}^{(k)}$ is the optimal $\boldsymbol{\hat \theta}_{\boldsymbol{\hat{\psi}}}$. 

\subsection{Posterior predictive inference with Monte-Carlo sampling}
 Given $\boldsymbol{\hat{\psi}}$ and the corresponding maximum a posteriori estimator $\boldsymbol{\hat{\theta}_{\hat{\psi}}}$ obtained during training, the predictive probability that a new test point $\boldsymbol{\tilde{x}}_i$ belongs to class $c=1, \ldots, K$ is
\begin{eqnarray*}
 p(\tilde{y}_i =c \mid \boldsymbol{\tilde{x}}_i; \boldsymbol{y}, \boldsymbol{x}) &=& \text{E}_{\boldsymbol{\tilde \theta} \sim p(\boldsymbol{\theta} \mid \boldsymbol{y}; \boldsymbol{x}, \boldsymbol{\hat \psi})} \left \{p(\tilde{y}_i =c \mid \boldsymbol{\tilde \theta}; \boldsymbol{\tilde{x}}_i) \right \} \\
& \approx & \cfrac{1}{S}\sum_{s=1}^S \text{Softmax}\left\{\boldsymbol{\tilde u}_{c}^{(s)}(\boldsymbol{\tilde{x}}_i)\right \},
\end{eqnarray*}
where each of the Monte-Carlo samples is generated by 
\begin{eqnarray*}
\boldsymbol{\tilde u}_{c}^{(s)}(\boldsymbol{\tilde{x}}_i) &=& \hat{\Phi}(\boldsymbol{\tilde{x}}_i)\boldsymbol{\tilde \beta}_c^{(s)} + \hat{d}_c(\boldsymbol{\tilde{x}}_i) \tilde{\kappa}^{(s)}_{c} + \boldsymbol{\hat{v}}_{c}(\boldsymbol{\tilde{x}}_i) \boldsymbol{\tilde \gamma}^{(s)},
\end{eqnarray*}
using the eigencomposition $\boldsymbol{H}_{\rm P}(\boldsymbol{\hat{\theta}_{\hat{\psi}}};\boldsymbol{\hat{\psi}}) = \boldsymbol{Q}\boldsymbol{L} \boldsymbol{Q}^T$ and
$\boldsymbol{\tilde{\theta}}^{(s)} = \boldsymbol{\hat{\theta}_{\hat{\psi}}} + \boldsymbol{Q}\boldsymbol{L}^{-1/2} \boldsymbol{Q}^T \boldsymbol{w}^{(s)}$, where $\boldsymbol{w}^{(s)} \sim \mathcal{N}(\boldsymbol{0}, \boldsymbol{I})$. We then predict to the class with the highest probability
\begin{eqnarray*}
y^{*}_i & = & \arg \max_{c} p(\tilde{y}_i =c \mid \boldsymbol{\tilde{x}}_i; \boldsymbol{y}, \boldsymbol{x}).
\end{eqnarray*}
Overall, the empirical Bayes approach to training the hetSNGP model enhances flexibility while enabling data-driven estimation of hyperparameters, requiring only the spectral-normalization factor and network architecture to be tuned, similar to the minimal tuning required by the competitive baseline DDU. We now assess the performance of our approach on large-scale simulated data from LHC.  The synthetic two circles toy dataset, comparing auto-hetSNGP with hetSNGP can be found in Appendix \ref{appendix_example_syn}.

\section{Results for LHC simulated tracking data}\label{sec:results}

%\paragraph{LHC simulated tracking data.} 
The training set consists of 1'065'273 particles, including 475'362 in the stable region and 589'911 in the unstable region. Each particle is described by 157 features that represent
\begin{itemize}
\item A seed index, specifying which of the sixty realizations of the nonlinear magnetic field errors assigned to the magnets of the accelerator model is used; 
\item 8 parameters describing the phase-coordinates (in terms of radius and angle of polar coordinates), the beam index (specifying whether the clockwise or counter-clockwise beam is used), the value of the linear betatron tunes, the value of the linear chromaticities, the strength of the Landau octupole magnets;
\item 10 MAD-X optimal parameters, 
\item 72 beam optics parameters.
\item 7 parameters describing the dependence of the betatron frequency with amplitude, which are related to the nonlinear dynamics.
\end{itemize}

The test set comprises 131'593 particles, with 57'918 in the stable region and 73'675 in the unstable region, so the stable and unstable classes comprise 44\% and 55\% of the data, respectively. We trained a fully deterministic residual neural network and the auto-hetSNGP model introduced in Section~\ref{sec:modeling}, and compared their uncertainty estimates across three different methods: Monte Carlo Dropout \citep{mcdrop}, DDU and the empirical Bayes introduced in Section~\ref{sec:ebopt}. DE was infeasible due to computational constraints. We distributed the training across 16 compute nodes for 250 epochs using mini-batches of size 512. We used an initial learning rate of 0.0001 combined with a cosine annealing scheduler with warm restarts every 100 iterations. In the auto-hetSNGP model, we set the number of Monte Carlo samples to $S = 2048$ and used $K=R=2$. Note that in practice, we found $R = K$ sufficient for this binary classification task, despite the low-rank assumption $R\ll K$ that motivates the general formulation.
     
Table~\ref{tab:trained} summarizes the characteristics of the best performing architectures. Since Monte Carlo Dropout is activated only at inference time, we used the same underlying model that was trained for DDU. Nevertheless, we duplicated its specifications for clarity and ease of reference. The three methods used the residual neural network illustrated in Figure~\ref{fig:res}, which consists of two linear layers per basic block and a single linear layer per shortcut block. The number of units per layer and the number of basic blocks vary between configurations. Specifically, both MC Dropout and DDU used a network with seven basic blocks. The number of units per layer decreases progressively across blocks: both layers of the first block have 256 units, those of the second block have 128 units, and so on, with the seventh block containing 4 units. Similarly, auto-hetSNGP consists of six basic blocks, each with two linear layers whose units are specified per block. Although both architectures exhibit comparable model complexity, auto-hetSNGP converges in significantly fewer training epochs.

\begin{table}[htb]
\caption{Distinguishing specifications of the best performing residual models. The ``units per layer'' column encodes two pieces of information: the length of the list is the number of basic blocks, each comprising two linear layers, while the values in the list specify the number of units in each layer. The auto-hetSNGP represents the approach proposed in this paper.}
\label{tab:trained}
\centering
\begin{tabular}{lcccc} 
\toprule
    \textbf{Model} & \textbf{Best epoch} & \textbf{Dropout rate} & \textbf{SN factor} & \textbf{Units per layer}\\
    \midrule
    %No dropout & 250 &  & 0.0 & 1.0 & $[256, 128, 64, 32, 16, 8, 4]$\\
    MC Dropout & 114  & 0.1 & 2.0 & $[256, 128, 64, 32, 16, 8, 4]$\\
    DDU & 114 & 0.1 & 2.0 & $[256, 128, 64, 32, 16, 8, 4]$\\
    auto-hetSNGP & 39 & 0.1 & 1.1 & $[256, 128, 64, 32, 16, 8]$\\
    \bottomrule\\
\end{tabular}
\end{table}

Table~\ref{tab:best_perf} shows a performance comparison of the best models evaluated on the test set, using the Expected Calibration Error (ECE)~\citep{ece, guo2017calibration} to measure uncertainty calibration. Let $\mathrm{TP}$, $\mathrm{TN}$, $\mathrm{FP}$, and $\mathrm{FN}$ denote the number of true positives, true negatives, false positives, and false negatives, respectively, positive being the stable region and negative being the unstable one. The metrics used for evaluating accuracy are defined as
\begin{eqnarray*}
  \text{Sensitivity} &=& \frac{\mathrm{TP}}{\mathrm{TP} + \mathrm{FN}},  \quad \text{Precision}  = \frac{\mathrm{TP}}{\mathrm{TP} + \mathrm{FP}}, \quad
  \text{Specificity} = \frac{\mathrm{TN}}{\mathrm{TN} + \mathrm{FP}}, \\[4pt]
  F_1\text{-score} &=& 2 \cdot \frac{\text{Precision} \cdot \text{Sensitivity}}
                       {\text{Precision} + \text{Sensitivity}}.
\end{eqnarray*}
The ECE measure is a weighted average over equally spaced bins $B_1, \ldots, B_M$ of the data:
\begin{eqnarray*}
ECE &=& \sum_{m=1}^M \cfrac{|B_m|}{n} \left| \mathrm{acc}(B_m) - \mathrm{conf}(B_m)\right|, \\
\mathrm{acc}(B_m) &=& \cfrac{1}{|B_m|}\sum_{i \in B_m}1\left\{y^{*}_i = y_i\right\}, \quad \mathrm{conf}(B_m) = \cfrac{1}{|B_m|}\sum_{i \in B_m}p(y^{*}_i \mid \boldsymbol{\tilde{x}}_i; \boldsymbol{y}, \boldsymbol{x}).
\end{eqnarray*}
The MC Dropout results were obtained by averaging the softmax outputs over 5 stochastic forward passes. 
\begin{table}[H]
\caption{Performance of the best models evaluated on the test set. The auto-hetSNGP represents the approach proposed in this paper.}
\label{tab:best_perf}
\centering
\begin{tabular}{lcccccccc}
\toprule
\textbf{Model} & \multicolumn{7}{c}{\textbf{Accuracy [\%]}} & {\textbf{Uncertainty}} \\
\cmidrule(lr){2-8} \cmidrule(lr){9-9}
               & Stable & Unstable & Overall & Sensitivity & Precision & Specificity & $F_1$-score & ECE \\
\midrule
MC Dropout     & 94.98 & 97.06 & 96.14 & 94.98 & 96.21 & 97.06 & 95.59 & 0.020\\
DDU            &    96.83 & 97.14 & 97.00 & 96.83 & 96.38 & 97.14 & 96.60 & 0.016 \\
auto-hetSNGP   & 96.34 & 97.19 & 96.82 & 96.34 & 96.42 & 97.19 & 96.38 & 0.019 \\
\bottomrule\\
\end{tabular}
\end{table}

Overall, DDU and auto-hetSNGP achieve comparable results on the test set, marginally outperforming MC Dropout. However, DDU's uncertainty estimates become less interpretable on the three test configurations originally shown in Figure~\ref{fig:original}. In Appendix \ref{appendix_extra_configs} The corresponding uncertainty visualizations are presented in Figures~\ref{fig:1186}--\ref{fig:29876}, where the high uncertainty is depicted in red and the low uncertainty in yellow. For both MC Dropout and auto-hetSNGP, we use the predictive variance, rescaled by a factor of 0.25 to be within the interval $(0, 1)$, as a proxy for uncertainty. 

The ideal behavior for uncertainty estimation is that the model exhibits high uncertainty when making incorrect predictions and low uncertainty when the predictions are correct. In particular, the decision boundary, where class overlap and ambiguity are expected, should correspond to regions of high uncertainty, while the interior of well-separated classes should exhibit low uncertainty. However, DDU displays counterintuitive behavior: for configurations 1186 and 29876, it shows higher uncertainty in the stable region than in the unstable region, and its uncertainty estimates are unreliable for configuration 29472. In contrast, auto-hetSNGP produces more consistent and interpretable uncertainty estimates that align more closely with the expected behavior. Interestingly, these three configurations highlight the trade-off between robustness and flexibility that is often required in real-world applications. Although DDU achieves slightly better overall performance on the full test set, auto-hetSNGP outperforms it on localized test configurations, demonstrating that increased model flexibility can provide a tangible advantage in challenging scenarios. This difference stems from the underlying modeling assumptions: DDU fits a single, shared Gaussian distribution for each class, whereas auto-hetSNGP explicitly incorporates class-dependent noise, allowing it to better capture heteroscedastic uncertainty across regions of the input space. However, the overall poor performance of the MC dropout suggests that the model may be overparameterized, with important units being deactivated at inference time due to the stochastic nature of the dropout. We note that, due to computational constraints, the results reflect a single training run per method; variance between random seeds was not assessed.

\section{Discussion}\label{sec:discussion}
We considered the computational and modeling challenges associated with uncertainty quantification in large-scale, simulation-driven scientific problems. Specifically, we proposed a flexible hetSNGP model trained with an empirical Bayesian approach that integrates hyperparameter optimization directly into the training loop. This enables efficient and robust learning without requiring costly validation or manual tuning, two common bottlenecks in Bayesian neural models. Our approach extends the SNGP framework by incorporating class-dependent heteroscedasticity with learnable variances, allowing the model to represent both predictive uncertainty and structured noise in a principled manner. We demonstrated the practical benefits of this architecture on the task of DA prediction in circular particle accelerators, where accurate uncertainty estimation is critical for guiding expensive tracking simulations. Compared to existing methods, our approach achieves competitive predictive performance while offering well-calibrated uncertainty estimates and training efficiency similar to the simplest state-of-the-art method. In particular, it is more computationally tractable than DE and exhibits a more consistent uncertainty behavior than DDU on configuration-specific test cases, where standard methods tend to misrepresent class boundaries and overestimate confidence. Our technical contribution is the separation of the model’s parameters into deterministic components $\boldsymbol{\psi}$ and stochastic components $\boldsymbol{\theta}$, with distinct functions during training and inference. By restricting optimization to the deterministic parameters $\boldsymbol{\psi}$ of the neural network, we accelerate the training process. The stochastic parameters $\boldsymbol{\theta}$, which govern the predictive uncertainty of the model, are then sampled at inference time using the Laplace approximation centered on their mode. This separation enables scalable and memory-efficient training, while still allowing for a semi-Bayesian treatment of uncertainty at test time. The low dimensionality of $\boldsymbol{\theta}$ makes posterior inference tractable via a preconditioned Newton--Raphson procedure, further contributing to the method’s computational efficiency.

Our framework is broadly applicable beyond accelerator physics. Many domains, such as climate science, biological modeling, and engineering design, suffer from similar computational constraints and demand high-fidelity uncertainty estimates for decision-making. The empirical Bayes methodology we propose generalizes to these contexts, offering a path forward for scalable uncertainty-aware modeling under resource limitations. Future work could explore replacing the Laplace approximation with more complex posterior inference techniques. Finally, coupling our method with acquisition functions in active learning or Bayesian optimization pipelines may further reduce the need for expensive simulations by focusing computational effort on uncertain regions of parameter space.

\subsubsection*{Acknowledgments}
This work is funded by the Swiss Data Science Center project grant C20-10.

\bibliography{references}
\bibliographystyle{plainnat}

\appendix

\section{Performance on different configurations}
\label{appendix_extra_configs}
We now present different configurations corresponding to low, medium, and high DA. Intuitively, the boundary between the stable and unstable regions should correspond to higher uncertainty estimates, while the complementary regions are expected to exhibit lower uncertainty. For the predicted boundary, a well-performing method is expected to demonstrate both a low number of prediction errors around the DA region and a qualitatively reasonable concentration of uncertainty around the predicted boundary.

\subsection{Comparison for configuration 1186. Medium DA.}
\begin{figure}[H]
\begin{minipage}[H]{.49\textwidth}
\centering
\includegraphics[width=\textwidth]{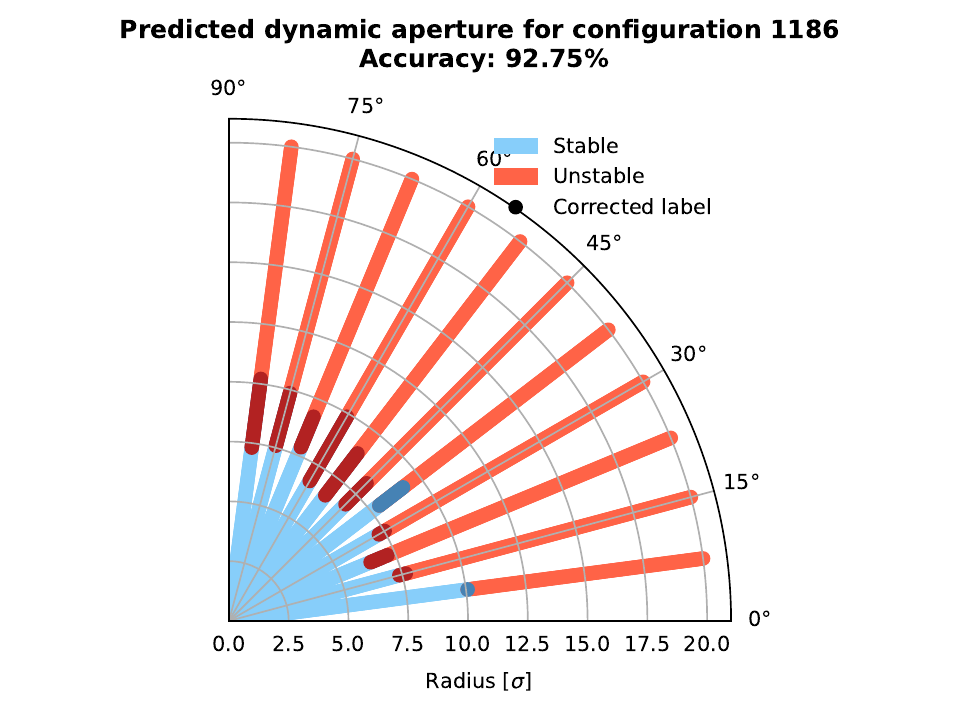}
\end{minipage}
\hfill
\begin{minipage}[H]{.49\textwidth}
\centering
\includegraphics[width=\textwidth]{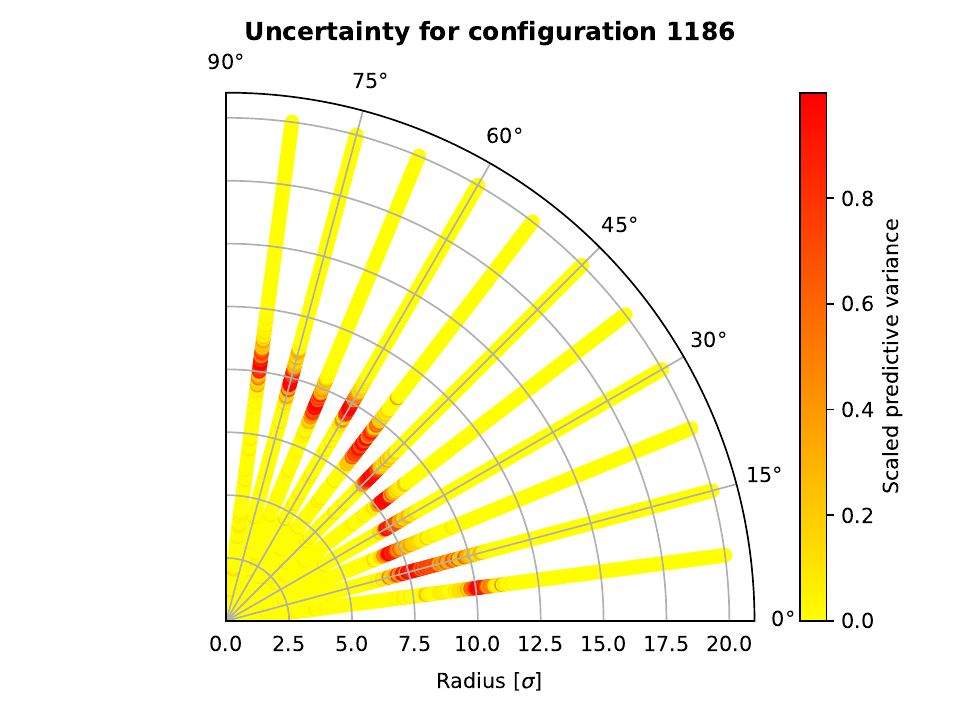}
\end{minipage} 
\centering
a) Performance of MC Dropout.
%\end{figure}
\vspace{2em}
\vfill
%\begin{figure}[H]
\begin{minipage}[H]{.49\textwidth}
\centering
\includegraphics[width=\textwidth]{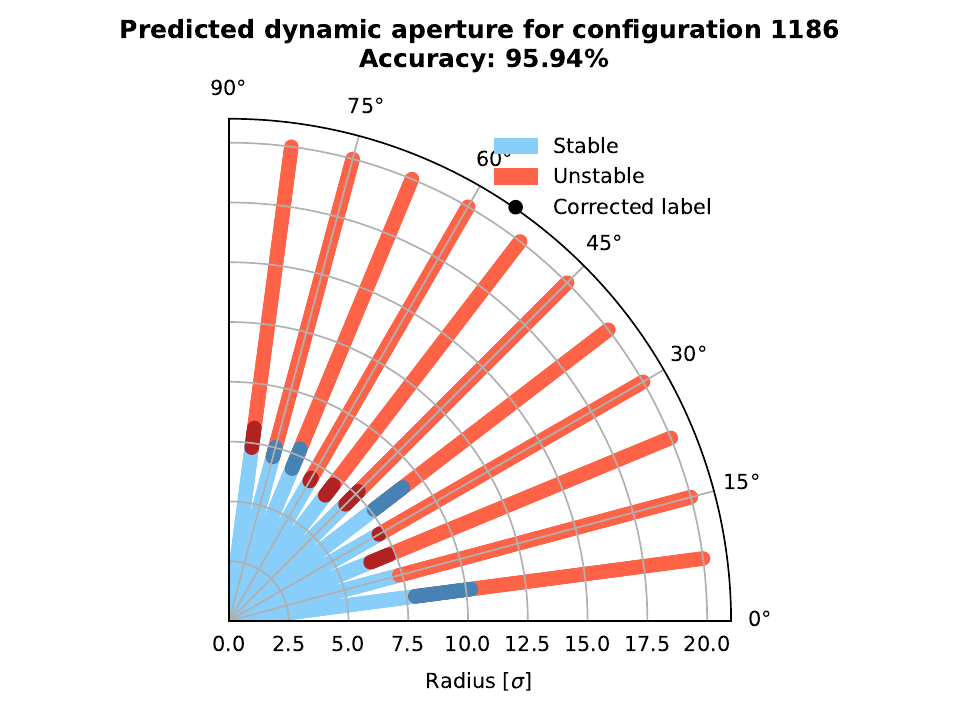}
\end{minipage}
\hfill
\begin{minipage}[H]{.49\textwidth}
\centering
\includegraphics[width=\textwidth]{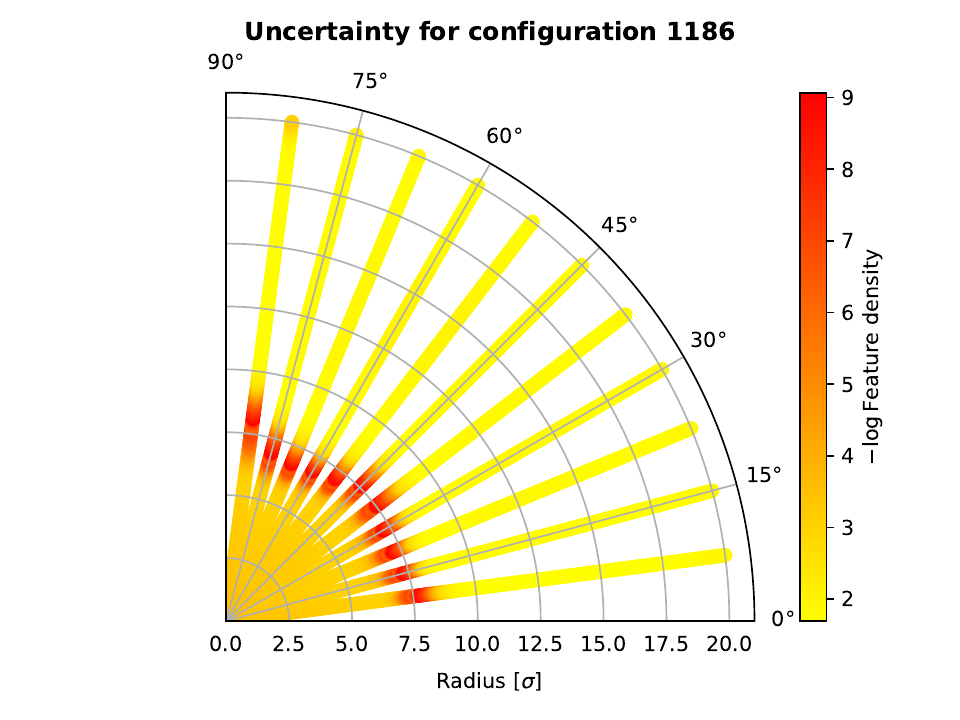}
\end{minipage} 
\centering
b) Performance of DDU.
%\end{figure}
\vspace{2em}
\vfill
%\begin{figure}[H]
\begin{minipage}[H]{.49\textwidth}
\centering
\includegraphics[width=\textwidth]{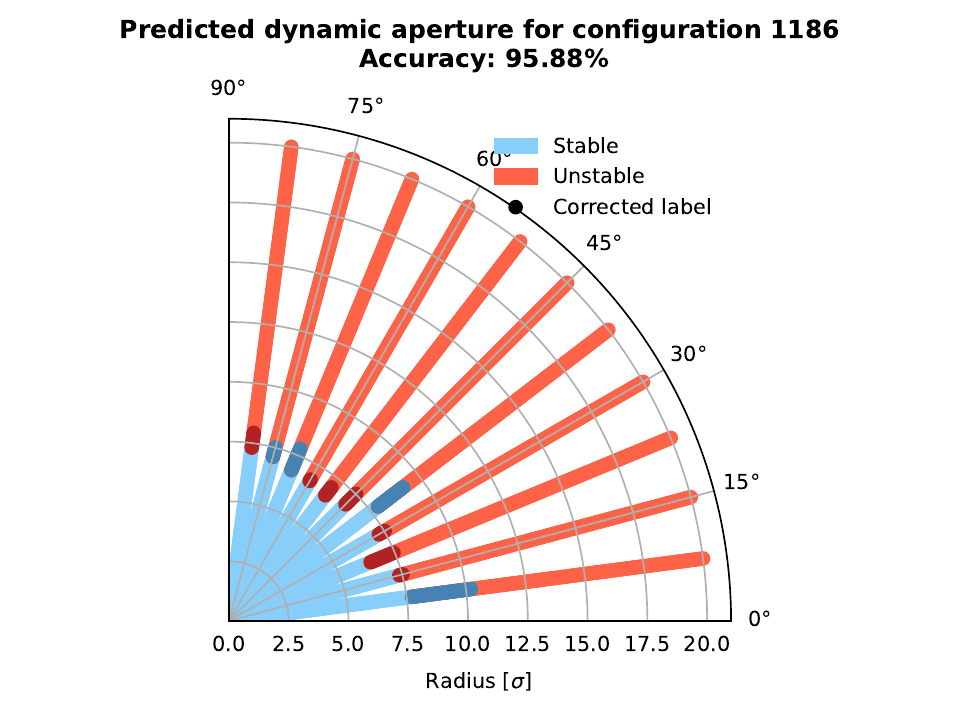}
\end{minipage}
\hfill
\begin{minipage}[H]{.49\textwidth}
\centering
\includegraphics[width=\textwidth]{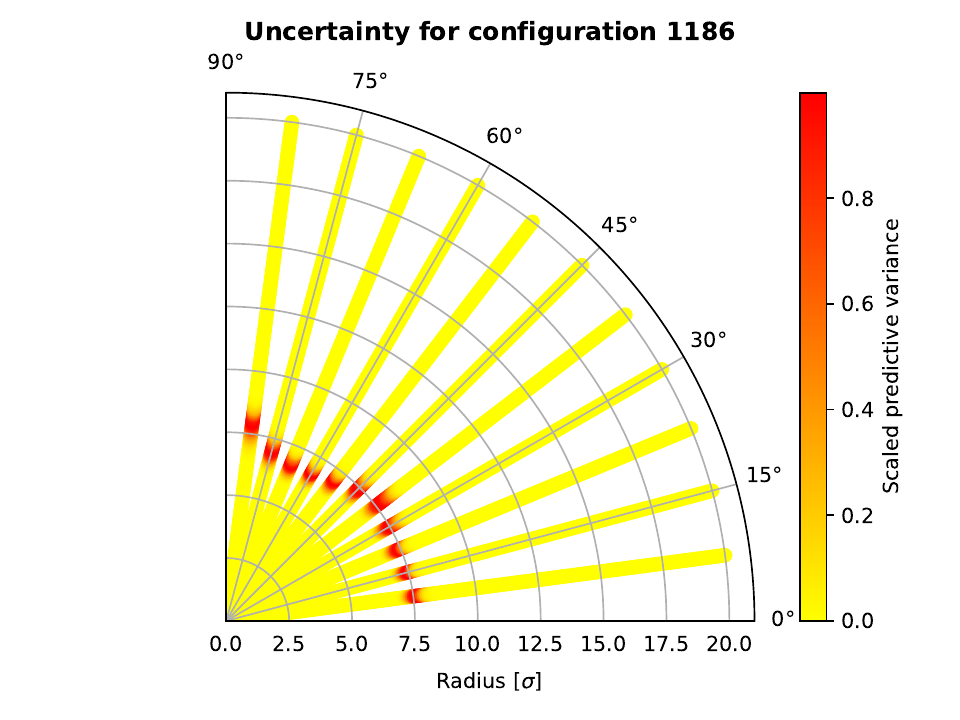}
\end{minipage} 
\centering
c) Performance of auto-hetSNGP.
\caption{Comparison for configuration 1186. On the left: predicted stability of the particles. If the model prediction was not correct, the true label is  depicted with darker color. On the right: uncertainty for the prediction.}\label{fig:1186}
\end{figure}

In Figure \ref{fig:1186}, the plots on the left show that all methods perform reasonably well, although MC Dropout overestimates the DA for larger angles. Based on the plots on the right, the uncertainty values qualitatively correspond to the expected behavior for MC Dropout and auto-hetSNGP, but not for DDU, for which the uncertainty is high in the stable-particle region.

\subsection{Comparison for configuration 29472. Low DA.}
\begin{figure}[H]
\begin{minipage}[H]{.49\textwidth}
\centering
\includegraphics[width=\textwidth]{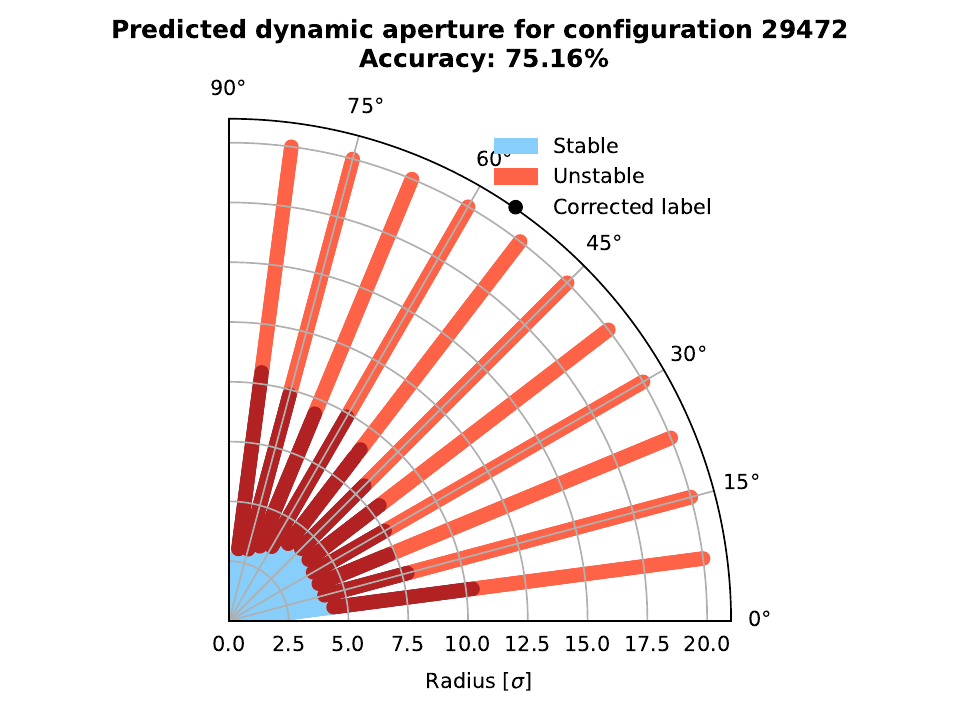}
\end{minipage}
\hfill
\begin{minipage}[H]{.49\textwidth}
\centering
\includegraphics[width=\textwidth]{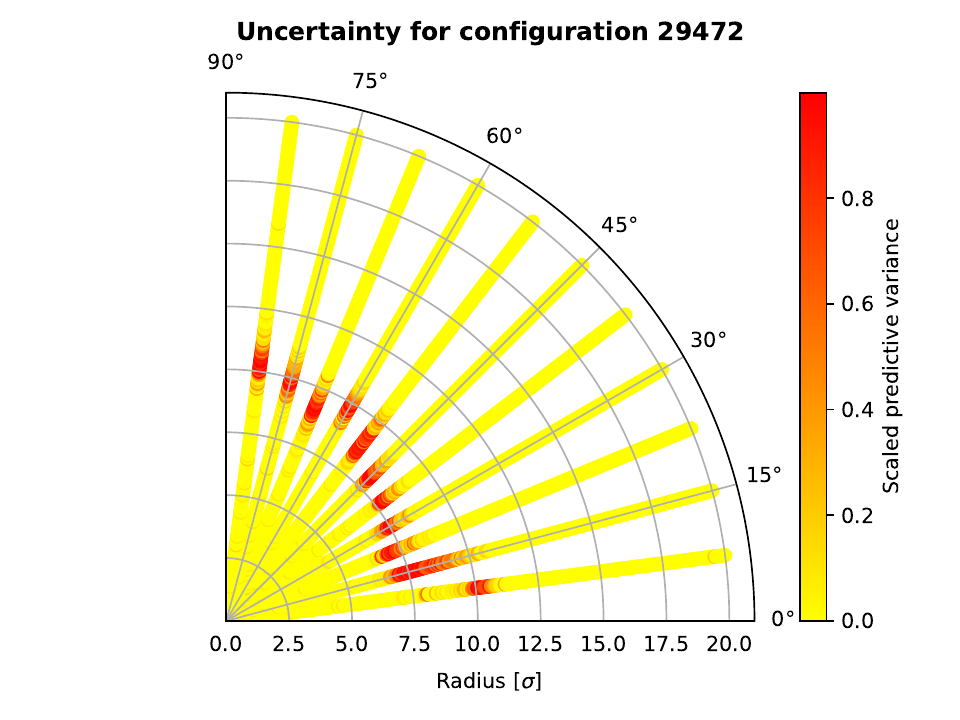}
\end{minipage} 
\centering
a) Performance of MC Dropout.
%\end{figure}
\vspace{2em}
\vfill
%\begin{figure}[H]
\begin{minipage}[H]{.49\textwidth}
\centering
\includegraphics[width=\textwidth]{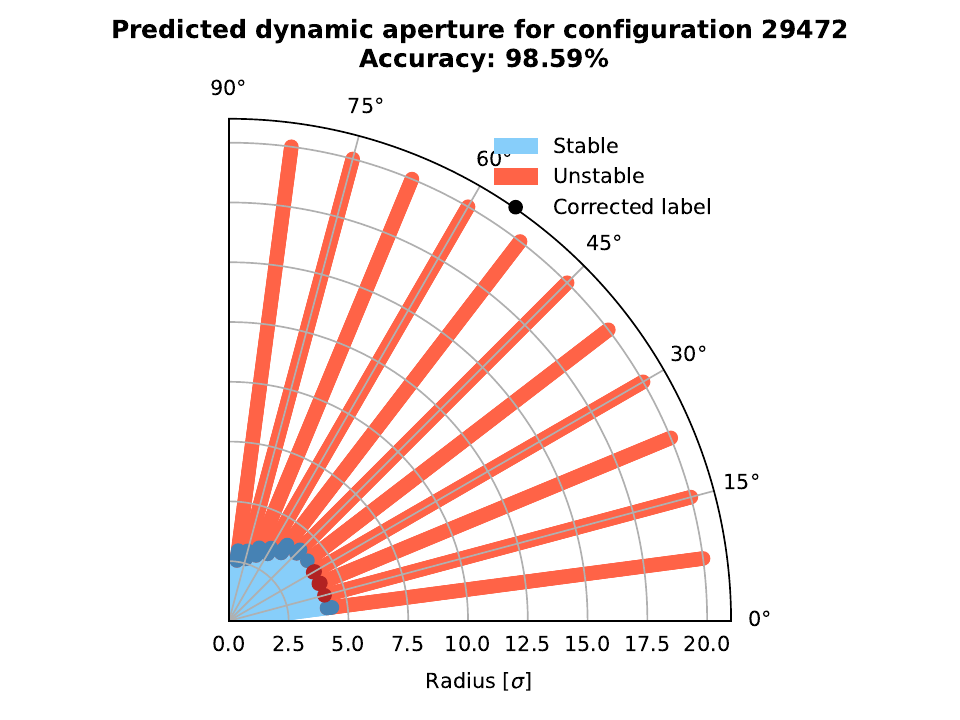}
\end{minipage}
\hfill
\begin{minipage}[H]{.49\textwidth}
\centering
\includegraphics[width=\textwidth]{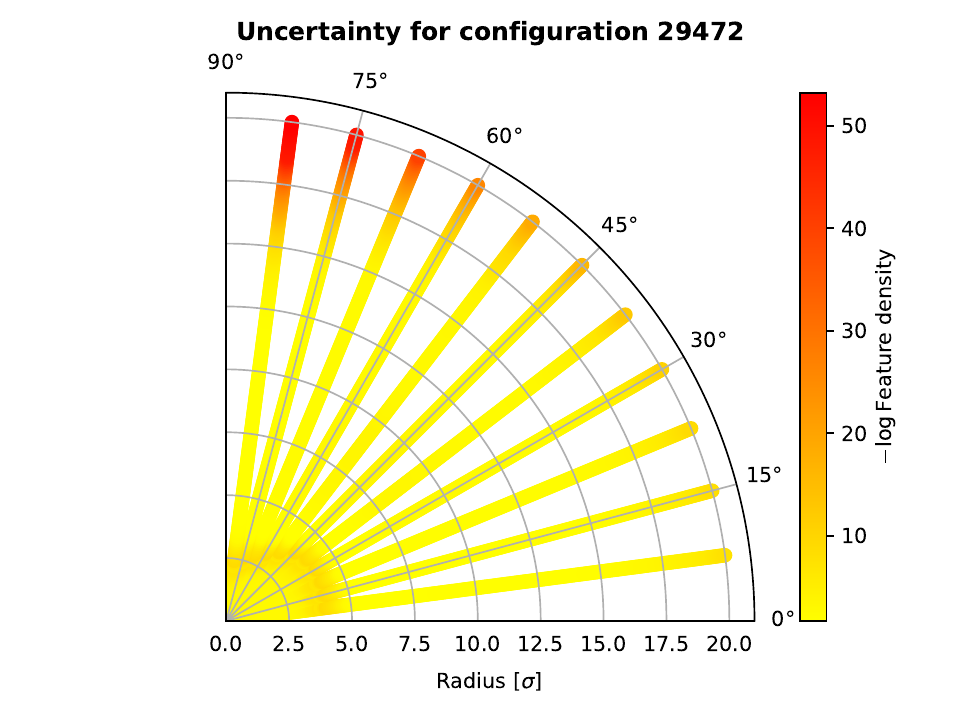}
\end{minipage} 
\centering
b) Performance of DDU.
%\end{figure}
\vspace{2em}
\vfill
%\begin{figure}[H]
\begin{minipage}[H]{.49\textwidth}
\centering
\includegraphics[width=\textwidth]{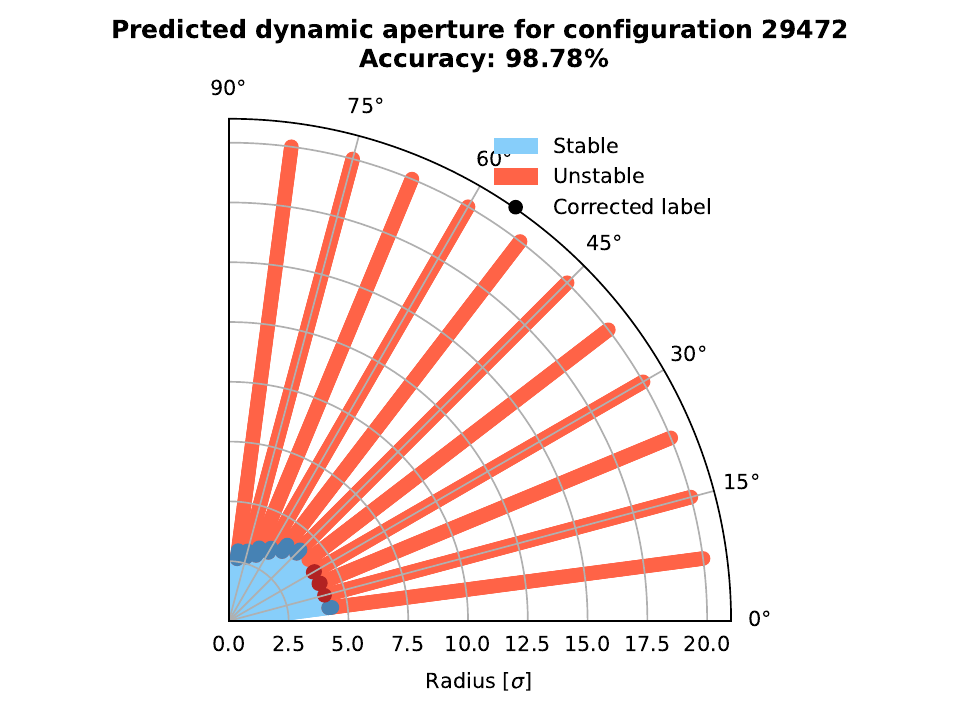}
\end{minipage}
\hfill
\begin{minipage}[H]{.49\textwidth}
\centering
\includegraphics[width=\textwidth]{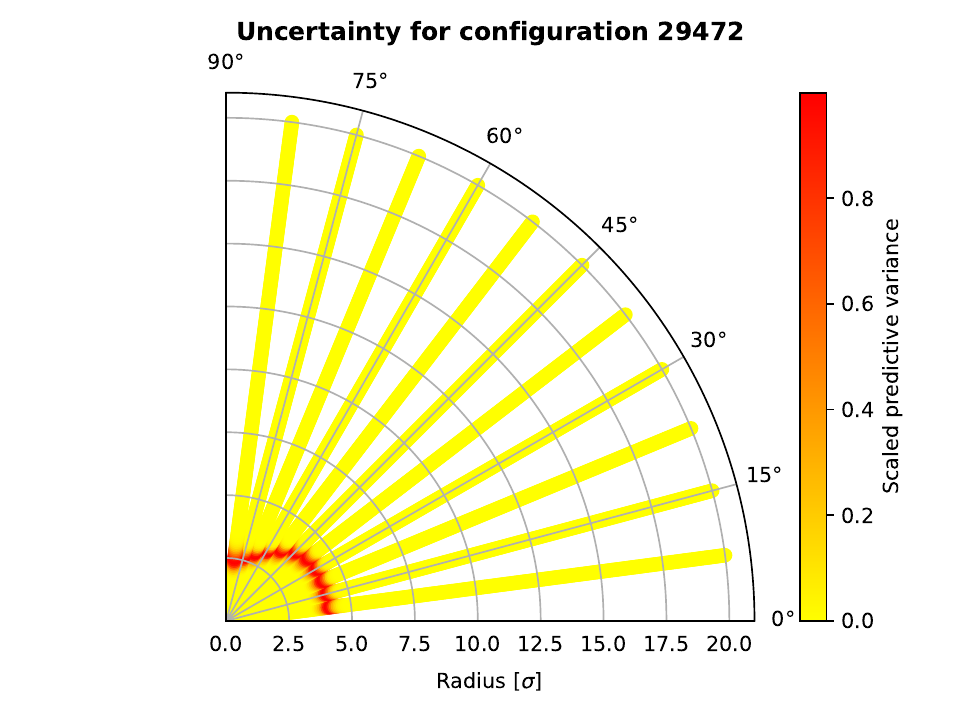}
\end{minipage} 
\centering
c) Performance of auto-hetSNGP.
\caption{Comparison for configuration 29472. On the left: predicted stability of the particles. If the model prediction was not correct, the true label is  depicted with darker color. On the right: uncertainty for the prediction.}\label{fig:29472}
\end{figure}

In Figure \ref{fig:29472}, based on the plots on the left, DDU and auto-hetSNGP demonstrate good overall accuracy, whereas MC Dropout significantly overestimates the DA. Based on the plots on the right, MC Dropout uncertainty estimates are high, as expected, around the incorrectly predicted class boundary. DDU demonstrates unexpectedly high uncertainty in the unstable region, while auto-hetSNGP reasonably highlights the uncertain boundary.

\subsection{Comparison for configuration 29876. High DA.}
\begin{figure}[H]
\begin{minipage}[H]{.49\textwidth}
\centering
\includegraphics[width=\textwidth]{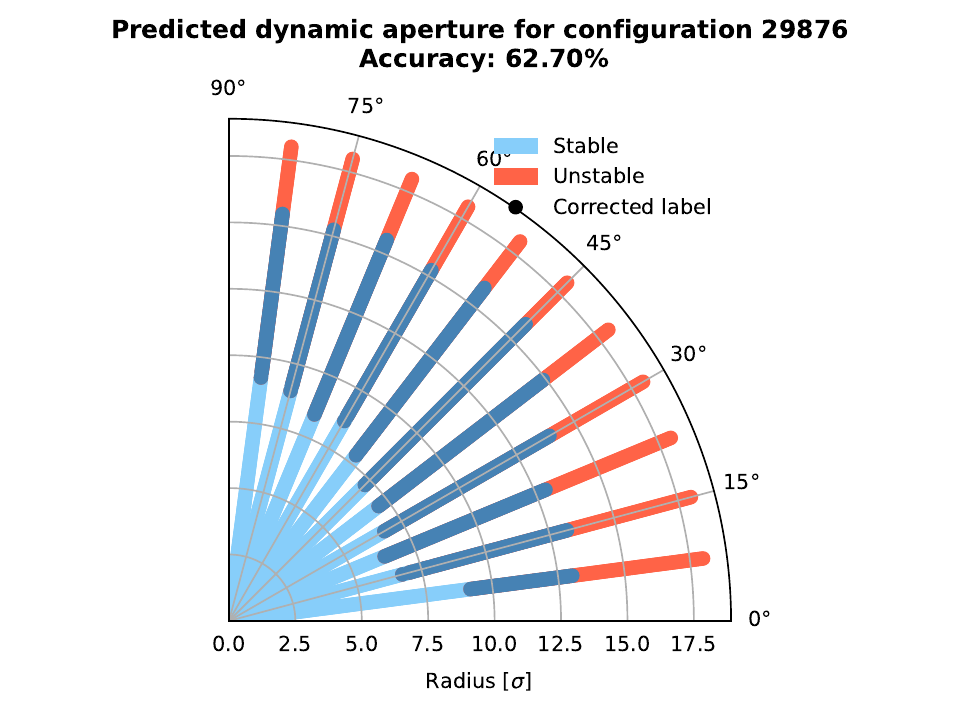}
\end{minipage}
\hfill
\begin{minipage}[H]{.49\textwidth}
\centering
\includegraphics[width=\textwidth]{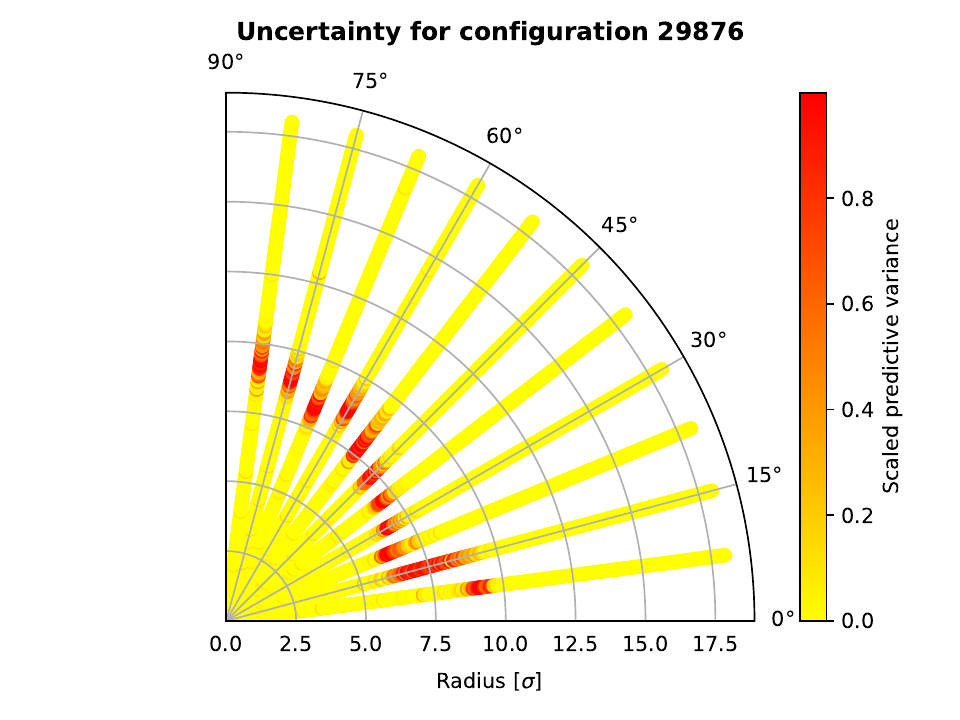}
\end{minipage} 
\centering
a) Performance of MC Dropout.
%\end{figure}
\vspace{2em}
\vfill
%\begin{figure}[H]
\begin{minipage}[H]{.49\textwidth}
\centering
\includegraphics[width=\textwidth]{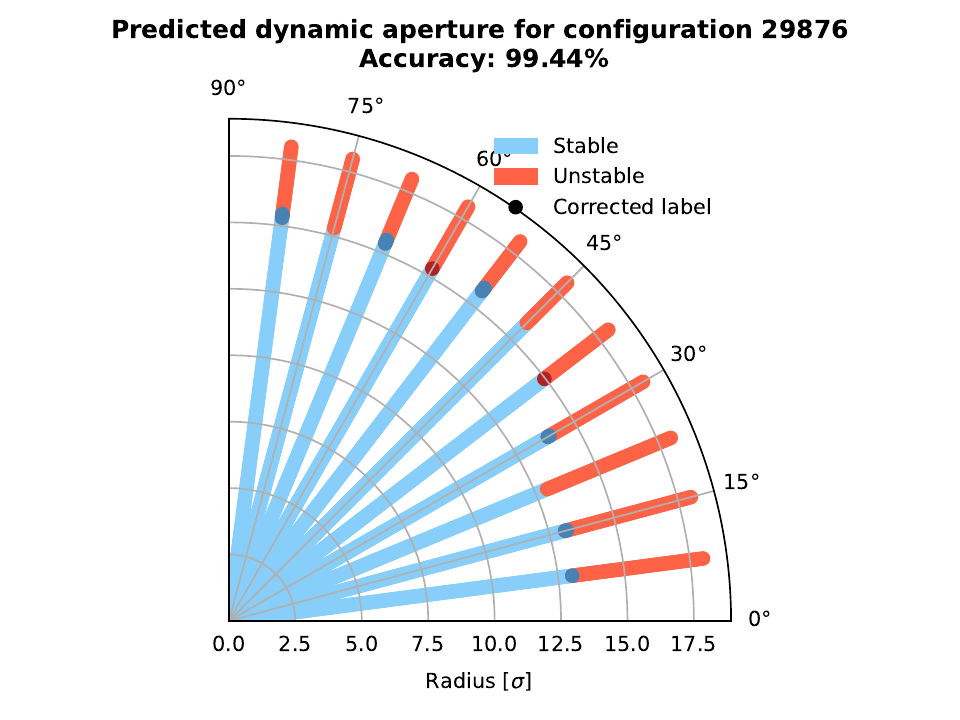}
\end{minipage}
\hfill
\begin{minipage}[H]{.49\textwidth}
\centering
\includegraphics[width=\textwidth]{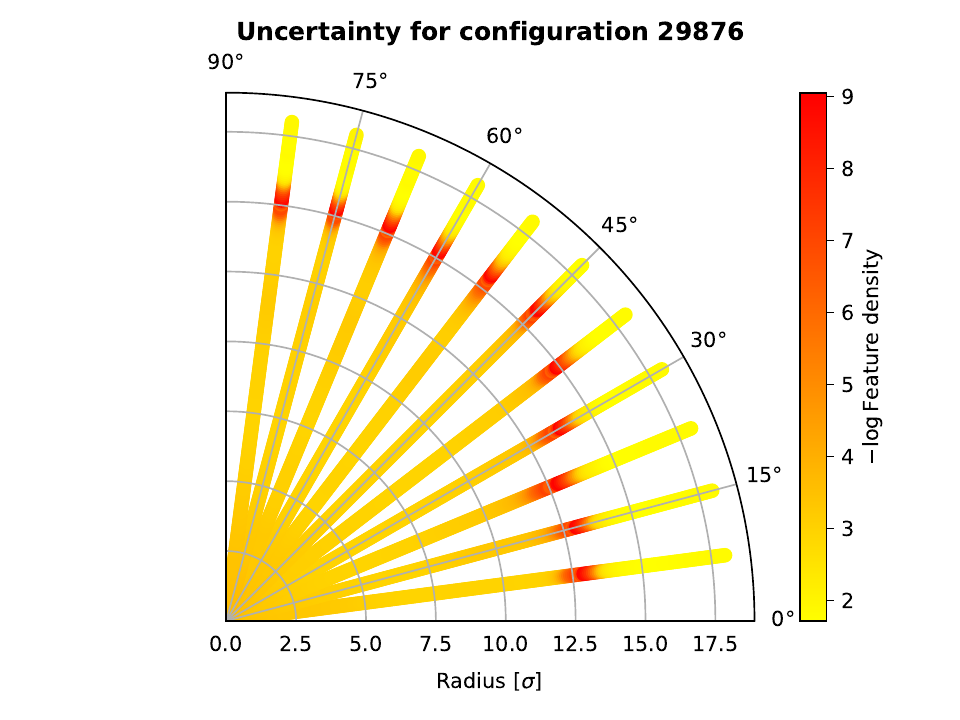}
\end{minipage} 
\centering
b) Performance of DDU.
%\end{figure}
\vspace{2em}
\vfill
%\begin{figure}[H]
\begin{minipage}[H]{.49\textwidth}
\centering
\includegraphics[width=\textwidth]{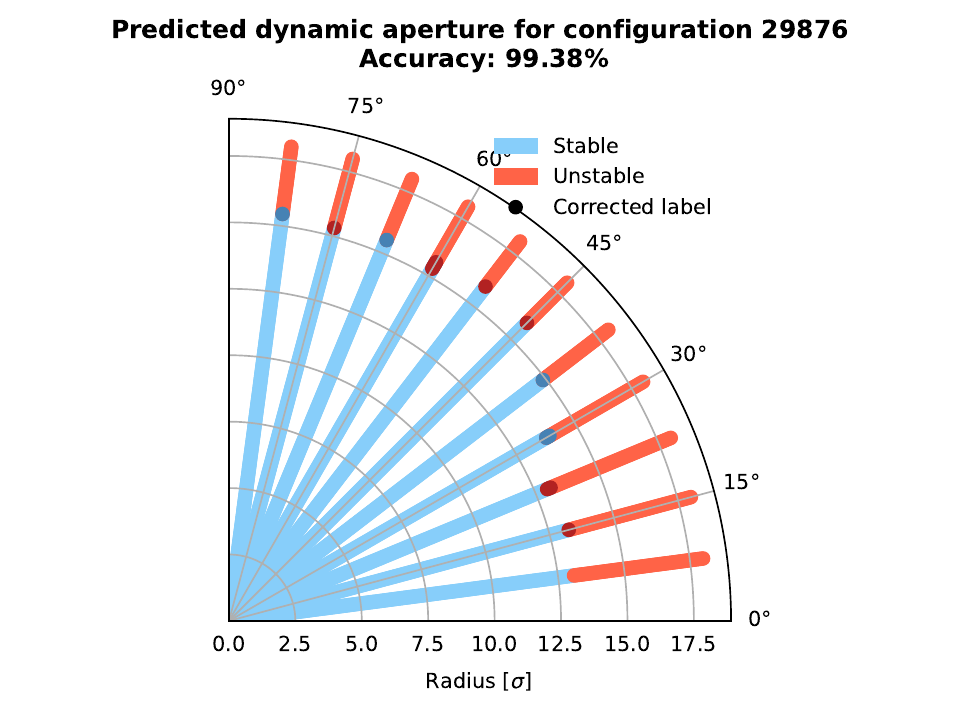}
\end{minipage}
\hfill
\begin{minipage}[H]{.49\textwidth}
\centering
\includegraphics[width=\textwidth]{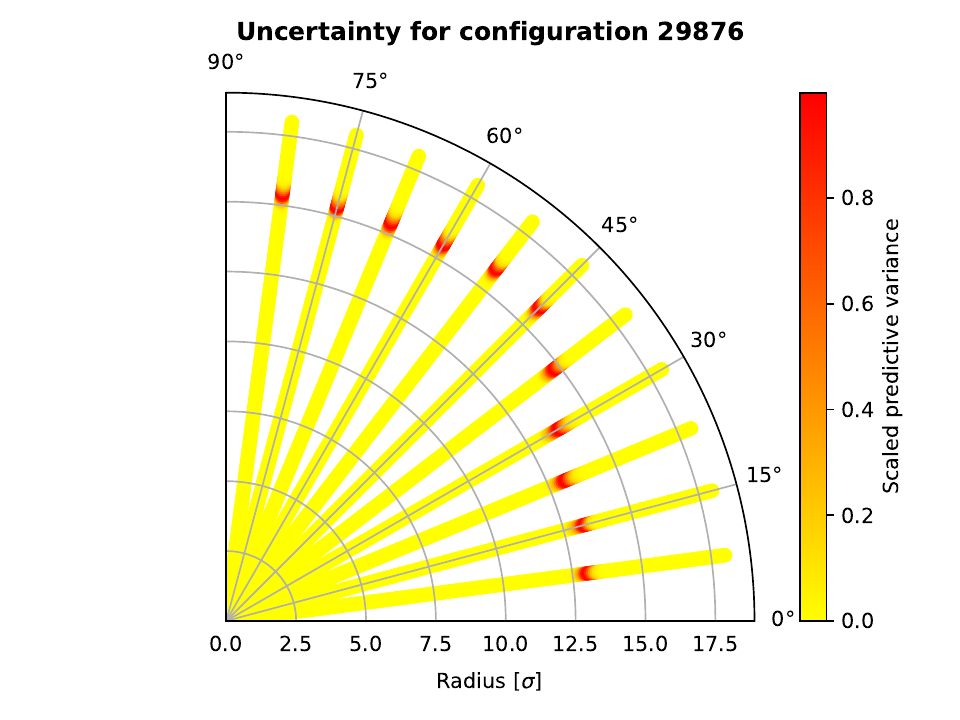}
\end{minipage} 
\centering
c) Performance of auto-hetSNGP.
\caption{Comparison for configuration 29876.}\label{fig:29876}
\end{figure}

In Figure \ref{fig:29876}, as in the previous configuration, the plots on the left show that DDU and auto-hetSNGP demonstrate good overall accuracy, whereas MC Dropout significantly underestimates DA. Based on the plots on the right, high MC Dropout uncertainty estimates are concentrated around the incorrectly predicted class boundary. DDU demonstrates unexpectedly high uncertainty in the stable region, while auto-hetSNGP reasonably highlights the uncertain boundary.
% config 1186

% config 29472

% config 29876

\section{2D example} 
\label{appendix_example_syn}
\begin{figure}[H]  
\includegraphics[width=1\linewidth]{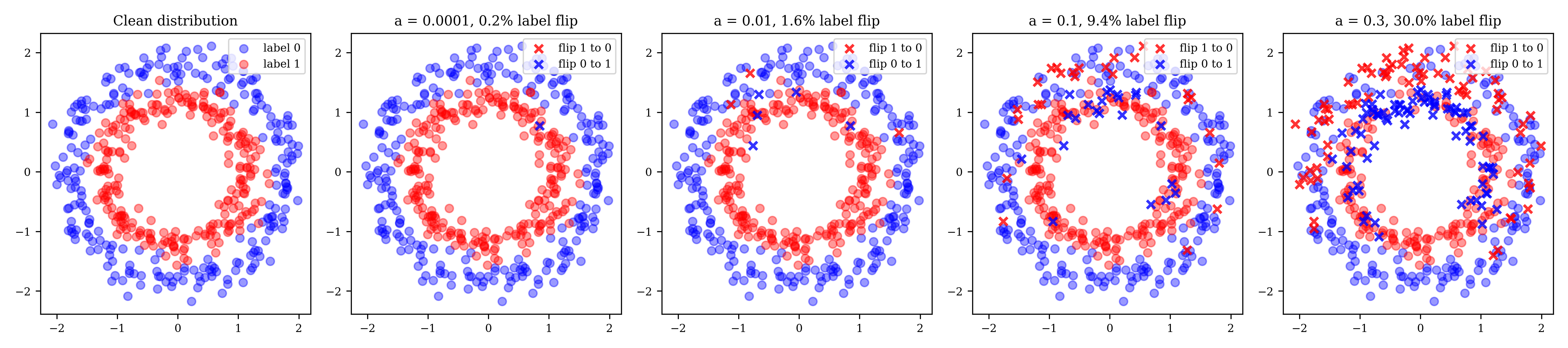} 
\caption{\label{two_circles} Data with different noise level contamination.}
\end{figure}

\paragraph{Two circles data.} We generate a synthetic binary classification dataset inspired by concentric circles with instance-dependent label noise. Let $\mathcal{D}_{\text{inner}}$ and $\mathcal{D}_{\text{outer}}$ denote uniform distributions along the circumference of an inner and an outer circle, respectively, each perturbed by isotropic Gaussian noise with variance $\sigma^2_{\text{geom}} = 0.1$. We sample the inputs $\boldsymbol{x} \in \mathbb{R}^2$ and the class label as follows
\begin{eqnarray*}
        y_{\text{true}} &=&
        \begin{cases}
            0, & \boldsymbol{x} \sim \mathcal{D}_{\text{inner}}, \\
            1, & \boldsymbol{x} \sim \mathcal{D}_{\text{outer}}.
        \end{cases}
\end{eqnarray*}
We flip each label with a probability that depends on the orientation of $\boldsymbol{x}$, $\boldsymbol{w}=(0,1)^\top$ say, such that 
\begin{eqnarray}
p_{\text{flip}}(\boldsymbol{x}) &=& a  \left(\frac{\boldsymbol{w}^\top \boldsymbol{x}}{\|\boldsymbol{w}\|\,\|\boldsymbol{x}\|}+1\right), \label{eq:flip}
\end{eqnarray}
where $a\in (0,1)$. The observed new label $\tilde{y}$ has then probability mass function $P(\tilde{y} = y_{\text{true}} \mid \boldsymbol{x}) = 1 - p_{\text{flip}}(\boldsymbol{x})$, with corresponding posterior
\begin{eqnarray*}
P(\tilde{y}=1 \mid \boldsymbol{x}) &=& P(\tilde{y} = y_{\text{true}} \mid \boldsymbol{x})P(y_{\text{true}}=1 \mid \boldsymbol{x}) + P(\tilde{y} \neq y_{\text{true}} \mid \boldsymbol{x})P(y_{\text{true}}=0 \mid \boldsymbol{x}) \\
&=& \left\{1-p_{\text{flip}}(\boldsymbol{x)}\right\}P(y_{\text{true}}=1 \mid \boldsymbol{x}) + p_{\text{flip}}(\boldsymbol{x})\left\{1-P(y_{\text{true}}=1 \mid \boldsymbol{x})\right\}.
\end{eqnarray*}
Figure \ref{two_circles} illustrates an example of such simulations for $a \in \{0.0001,0.01,0.1,0.3\}$ in  \ref{eq:flip}.
%We can now compare the variance of the perturbed labels given the input 
%    $ P(\tilde y=1 \mid x)\,\big(1 - P(\tilde y=1 \mid x)\big),
%    $
%and density of the data points ($ -\log p_{\mathcal D}(x)$), with various uncertainty estimates by the SNGP, Deep ensemble and DDU.
We now define the uncertainty measures used for evaluating the performance of hetSNGP, auto-hetSNGP, DE with 10 members, and DDU.  %Let $\hat{p}(\boldsymbol{x}) = \hat{P}(y = 1 \mid \boldsymbol{x})$ denote the predicted probability for class~1. 
Let $u(\boldsymbol{x})$ be  the second-to-last layer representation. 
We assess DE and SNGP-based models using the predictive variance of the logits %, scaled to the interval $(0,1)$,
\begin{equation}
\label{uncertainty_logit}
\hat U^{\text{var}}(\boldsymbol{x}) = {\rm{Var}}( u(\boldsymbol{x})),  
\end{equation}
where the expectation is taken with respect to the noise in posterior distribution for SNGP-type methods, or, for DE, with respect to the  randomness that defines ensemble members. 
For the DDU we use the logarithm of feature density estimated by DDU
\begin{equation}
\hat{U}^{\text{DDU}}(\boldsymbol{x}) = -\log p_{\text{DDU}} \left\{u(\boldsymbol{x})\right\}. \label{uncertainty_ddu}
\end{equation}

\begin{figure}[H] 
\centering
\includegraphics[width=\textwidth]{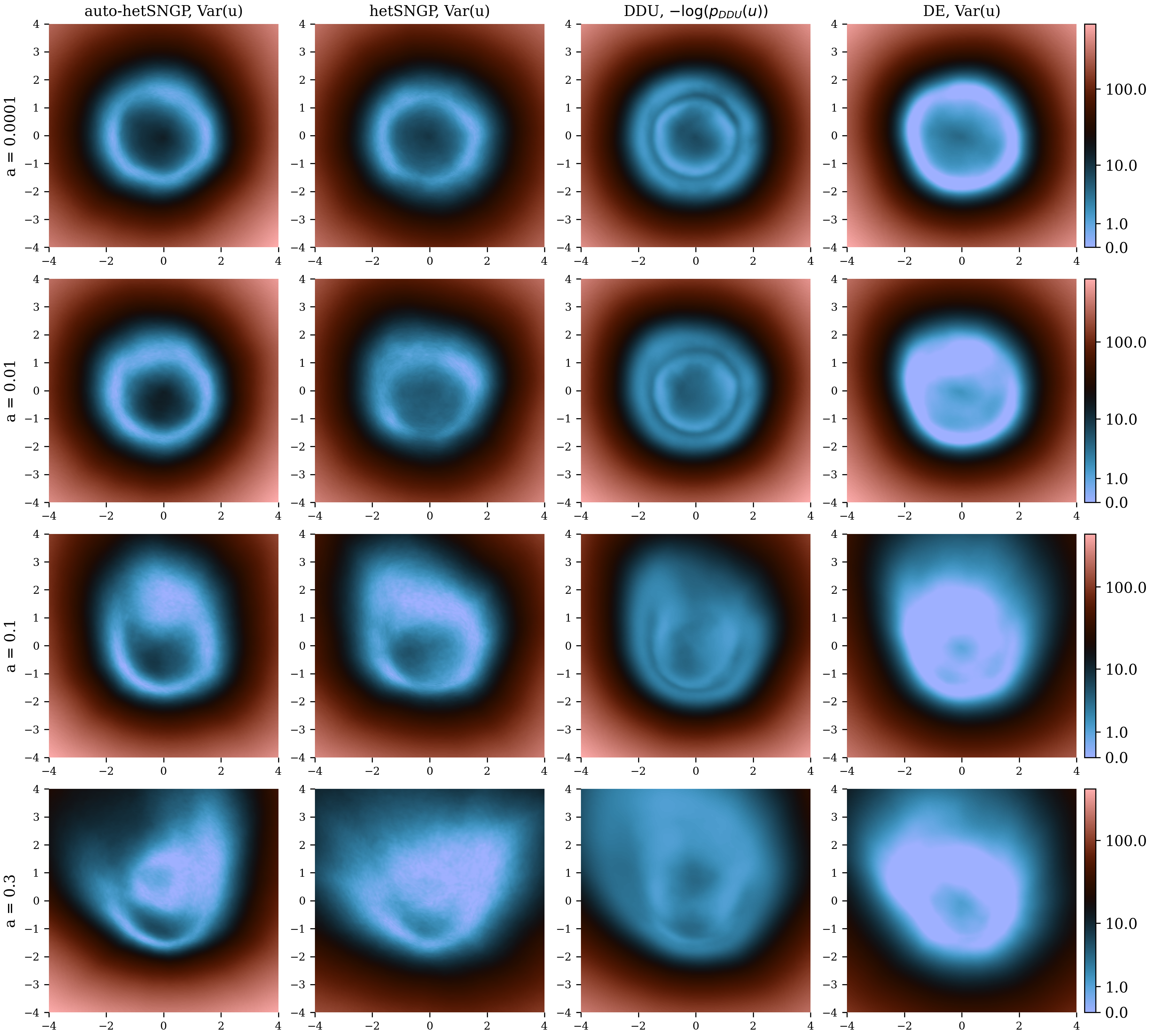} 
\caption{\label{two_circles_epi} Data density, \eqref{uncertainty_logit} variance of predictive logits based uncertainty estimates for hetSNGP methods and DE, DDU feature-density based uncertainty \eqref{uncertainty_ddu}.} 
\end{figure}

We randomly generated a training set of $1000$ points and a test set of 500 points, and illustrated the results for uncertainty quantification in Figure \ref{two_circles_epi}, where the lower uncertainty regions are highlighed with the blue color. All the models consisted of 4 ResNet blocks with width 128, dropout probability was set to 0.1. While the models exhibit comparable performance, auto-hetSNGP shows a marginally higher accuracy, see Table \ref{tab:accuracy}. Note that the DDU and DE baselines were not extensively tuned on this synthetic benchmark, so the accuracy gap relative to the (auto-)hetSNGP variants should not be over-interpreted as a general performance ranking.

\begin{table}[ht]
  \centering
  \caption{Test accuracy (\%) across methods and flip scales.}
  \label{tab:accuracy}
  \begin{tabular}{lcccc}
    \toprule
    Flip scale a &  auto-hetSNGP &  hetSNGP  & DDU & DE \\
    \midrule
    0.0001  & \textbf{92.0} & 90.0 & 75.6 & 74.0 \\
    0.01 & \textbf{91.2} & 90.0 & 74.0 & 74.4 \\
    0.1 & \textbf{86.4} & 80.0 & 67.6 & 68.4 \\
    0.3 & 62.8 & \textbf{63.6} & 57.2 & 58.8 \\
    \bottomrule
  \end{tabular}
\end{table}

%{\color{red}{ In \cite{}[add] the authors use the variance of the predicted logit as a proxy to (epistemic) uncertainty. This measure can be computed for SNGP-based models, and Deep ensemble from the logit $u_c(\boldsymbol{x})$ for class $c=1$  as 
%\begin{equation}
%    U^{\text{logit}}(\boldsymbol{x}) \;=\; 
%    \mathrm{Var}\!\left \{u_c(\boldsymbol{x})\right\}.
%     \label{uncertainty_logit}
%\end{equation}
%For the visualization purposes we rescale all the uncertainty estimates to the same scale ($[0,1]$). %{\color{red}{Why do DE and SNGP have different uncertainty measures?}}}}

%  {\color{red}{but we only see the results on the test set}} datasets: training and validation for tuning of the models {\color{red}{but hetSNGP is the only one that needs tuning}}, and a test dataset 

%\appendix
%\section{Appendix}
%You may include other additional sections here.

\end{document}